\documentclass[aps,preprint,preprintnumber,nofootinbib,amsmath,amssymb,bm,12pt]{revtex4}
\usepackage{graphicx,color}
\usepackage{bm,amsmath,amssymb}
\usepackage{ascmac}

\begin{document}

{
\small
\rightline{
\baselineskip16pt\rm\vbox to20pt{
\hbox{OCU-PHYS-536}
\hbox{AP-GR-167}
} } } 

\author{ {\large Ken Matsuno} }
\email{matsuno@sci.osaka-cu.ac.jp}

\affiliation{
Department of Mathematics and Physics, Graduate School of Science, Osaka City University,
Sumiyoshi, Osaka 558-8585, Japan
\bigskip}

\vskip 1cm

\title{ {\Large 
Hawking radiation of scalar particles and fermions 
from squashed Kaluza-Klein black holes 
based on a generalized uncertainty principle 
} \\ }

\begin{abstract}
We study the Hawking radiation from the five-dimensional charged static squashed Kaluza-Klein black hole 
by the tunneling of charged scalar particles and charged fermions. 
In contrast to the previous studies of Hawking radiation from squashed Kaluza-Klein black holes, 
we consider the phenomenological quantum gravity effects predicted by 
the generalized uncertainty principle with the minimal measurable length.  
We derive corrections of the Hawking temperature to general relativity,  
which are related to the energy of the emitted particle, the size of the compact extra dimension, 
the charge of the black hole and the existence of the minimal length in the squashed Kaluza-Klein geometry. 
We obtain some known Hawking temperatures in five and four-dimensional black hole spacetimes 
by taking limits in the modified temperature. 
We show that the generalized uncertainty principle 
may slow down the increase of the Hawking temperature due to the radiation, 
which may lead to the thermodynamic stable remnant of the order of the Planck mass 
after the evaporation of the squashed Kaluza-Klein black hole.   
We also find that the sparsity of the Hawking radiation 
modified by the generalized uncertainty principle 
may become infinite when the mass of the squashed Kaluza-Klein black hole approaches its remnant mass.      
\end{abstract}

\maketitle

\section{Introduction}
\label{sec:intro}

Hawking radiation is one of the interesting phenomena 
where both of general relativity and quantum theory play a role.    
While a black hole does not emit radiation from a classical point of view, 
it is shown that, using the semiclassical method of quantum theory in curved spacetimes,  
a black hole can radiate from an event horizon like a blackbody at the Hawking temperature 
which is proportional to the surface gravity of the black hole  
\cite{Hawking:1974sw}.  
Hawking radiation and its black hole temperature are derived by several approaches. 
Parikh and Wilczek proposed an elegant derivation of Hawking radiation on the basis of the tunneling method 
\cite{Parikh:1999mf}. 
The essential idea of the tunneling mechanism is that 
particle-antiparticle pairs are formed close to the horizon inside a black hole.  
While the ingoing particles moving toward the center of the black hole are trapped inside the horizon, 
a part of the outgoing particles escapes outside the horizon by the quantum tunneling effect. 
If the particle which comes out to our Universe has positive energy, 
we can regard such a particle outside the horizon as the radiation from the black hole.  
The Hawking temperature is derived by 
comparing the tunneling probability of an outgoing particle with the Boltzmann factor in thermodynamics.   
The tunneling method has also been used to obtain the Hawking radiation of scalar particles and fermions  
across an event horizon in some black hole geometries   
\cite{Srinivasan:1998ty, Angheben:2005rm, Kerner:2007rr, Kerner:2008qv}. 
Then the derivation of Hawking radiation on the basis of the tunneling mechanism 
has been actively discussed in the literature 
(see the reference \cite{Vanzo:2011wq} as a review).

A generalized uncertainty principle with a minimal measurable length 
is attracted a lot of interest to consider phenomenological quantum gravity effects.   
Though the debates on the quantum nature of black holes are crucial and long standing, 
there is no systematic study by quantum gravity so far 
to lay a solid theoretical foundation for the arguments.
Then some different models of quantum gravity are effective ways to understand gravity behaviors 
at a sufficiently small scale. 
The combination of relativistic and quantum effects implies 
that the conventional notion of distance would break down the latest at the Planck scale. 
The basic argument is that the resolution of small distances requires test particles of short wavelengths  
and thus of high energies.  
At such small scales,  
the gravitational effects by the high energies of test particles would significantly disturb  
the spacetime structure which was tried. 
Then some quantum gravity theories suggest that there would exist 
a finite limit to the possible resolution of distances, which would be of order the Planck length.  
From several studies in string theory, 
such a minimal measurable length would be obtained by a generalized uncertainty principle,  
which is a quantum gravity inspired modification to the conventional Heisenberg uncertainty principle.   
Then various types of generalized uncertainty principles have been heuristically derived from thought experiments 
and are often taken as phenomenological models that would accommodate a minimal length
\cite{Kempf:1994su, Kempf:1996nk, Hossenfelder:2012jw},  
though there exist attempts to make its formulations more mathematically rigorous  
\cite{Isi:2013cxa}. 
Generalized uncertainty principles have been applied to some different systems  
and played an important role to consider its corrections by supposed quantum gravity theories 
\cite{Adler:2001vs, Hossenfelder:2003jz, Medved:2004yu, Das:2008kaa, Tawfik:2014zca, Petruzziello:2020wkd}.    
Motivated by these discussions of generalized uncertainty principles,  
we are interested in performing in-depth studies of the quantum features of 
higher-dimensional black hole solutions.  
Quantum gravity effects predicted by generalized uncertainty principles   
on the Hawking radiation have been studied by the tunneling of various particles in a variety of spacetimes 
including vacuum, electrovacuum and with a vast array of scalar fields or effective fluids 
\cite{Chen:2013pra, Chen:2013ssa, Chen:2013tha, Chen:2014xgj, Feng:2015jlj, Casadio:2017sze, Kuntz:2019gka}.  
In this paper, we focus on the quantum tunneling radiation of charged particles 
coming from higher-dimensional black holes based on a generalized uncertainty principle 
in the spacetime with compact extra dimensions.

Higher-dimensional black hole solutions are actively discussed 
in the context of string theories and braneworld models.    
Since our observable world is effectively four dimensional,   
we can regard higher-dimensional black hole solutions with compactified extra dimensions 
as candidates of realistic models.  
We call these Kaluza-Klein black holes. 
The four-dimensional Schwarzschild metric uniquely describes 
the general relativistic gravitational field in vacuum with spherical symmetry.  
However, even if we impose asymptotic flatness to the four-dimensional part of 
the higher-dimensional spacetime model with Kaluza-Klein structure,   
the metric is not uniquely determined. 
A family of five-dimensional squashed Kaluza-Klein black hole solutions 
\cite{Dobiasch:1981vh, Gibbons:1985ac, Ishihara:2005dp, Stelea:2008tt, Tomizawa:2012nk} 
represent fully five-dimensional black holes near the squashed S$^3$ horizons 
and asymptote to effective four-dimensional spacetimes with a twisted S$^1$ 
as an extra dimension at infinity.  
Then we can regard a series of squashed Kaluza-Klein black hole solutions 
with a twisted compactified extra dimension 
as one of realistic higher-dimensional black hole models.   
Several aspects of squashed Kaluza-Klein black holes have been discussed, for example, 
thermodynamics 
\cite{Cai:2006td, Kurita:2007hu, Kurita:2008mj}, 
Hawking radiation 
\cite{Ishihara:2007ni, Matsuno:2011ca, Li:2011zzm, Stetsko:2014dda}, 
stabilities \cite{Kimura:2007cr, Kimura:2018whv}, 
gyroscope precession \cite{Matsuno:2009nz, Azreg-Ainou:2019ylk}, 
thin accretion disk \cite{Chen:2011wb},  
X-ray reflection spectroscopy \cite{Zhu:2020cfn}, 
light deflection \cite{Matsuno:2020kju}, 
strong gravitational lensing \cite{Liu:2010wh, Chen:2011ef, Sadeghi:2012bj, Sadeghi:2013ssa, Ji:2013xua} 
and black hole shadow \cite{Long:2019nox, 1828181}.

In this paper, we investigate the Hawking radiation by the tunneling of charged particles 
and its quantum gravity effects based on the generalized uncertainty principle 
in the five-dimensional charged static squashed Kaluza-Klein black hole spacetime.         
To the best our knowledge, Hawking radiation as a tunneling process 
in the presence of a generalized uncertainty principle 
has not been discussed in asymptotically Kaluza-Klein spacetimes.  
Then, in contrast to the previous studies of Hawking radiation from squashed Kaluza-Klein black holes 
\cite{Ishihara:2007ni, Matsuno:2011ca, Li:2011zzm, Stetsko:2014dda}, 
we extend the derivation of Hawking radiation 
on the basis of the tunneling mechanism in four-dimensional black hole spacetimes    
to the case of the five-dimensional squashed Kaluza-Klein black hole,  
including the phenomenological quantum gravity effects predicted by 
the generalized uncertainty principle with the minimal measurable length.

According to the discussion of the Hawking radiation in general relativity,    
primordial mini black holes in the very early Universe would have been completely evaporated  
\cite{Sasaki:2018dmp}.   
However, during the final stages of the Hawking evaporation, 
the semiclassical approach would be expected to break down 
due to the dominance of quantum gravity effects.    
If the black holes would not evaporate completely 
but turn into the remnants at the end of the evaporation process,    
such remnants might be a constituent of dark matter,  
similar to other candidates including weakly interacting massive particles, 
the axion, the axino, the neutralino and the gravitino 
\cite{MacGibbon:1987my, Chen:2002tu, Chen:2004ft, 
Scardigli:2010gm, Dalianis:2019asr, Lehmann:2019zgt, Bai:2019zcd}.  
Then it is interesting to investigate how higher-dimensional black holes with compact extra dimensions 
evolve under a Hawking evaporation process modified by a generalized uncertainty principle. 
In this paper, we study the evaporation of the squashed Kaluza-Klein black hole 
by the tunneling of particles in the framework of the generalized uncertainty principle 
as one of the quantum gravity effects in the Hawking radiation.

This paper is organized as follows. 
In the section \ref{sec:kkbh}, 
we review the properties of five-dimensional charged static Kaluza-Klein black hole solutions 
with squashed horizons. 
In the section \ref{sec:stunneling}, we consider the tunneling radiation of charged scalar particles 
from the squashed Kaluza-Klein black hole in the context of the generalized uncertainty principle 
and derive the modified Hawking temperature with the quantum gravity effects.    
We obtain some known Hawking temperatures in five and four-dimensional black hole spacetimes 
by taking limits in the modified temperature. 
Then we investigate the evaporation process of the black hole and the sparsity of the radiation 
by the tunneling of scalar particles in the squashed Kaluza-Klein geometry.  
In the section \ref{sec:ftunneling}, we study the Hawking radiation by the tunneling of charged fermions 
from the squashed Kaluza-Klein black hole based on the generalized uncertainty principle.   
The section \ref{sec:summary} is devoted to summary and discussion.

\section{Squashed Kaluza-Klein black holes}
\label{sec:kkbh}

We consider the charged static Kaluza-Klein black holes with squashed S$^3$ horizons 
in the five-dimensional Einstein-Maxwell theory 
\cite{Ishihara:2005dp}. 
The metric and the Maxwell field are respectively given by 
\begin{align}
 & ds^2 = -F dt^2 + \frac{K^2 }{F} d\rho^2 + \rho^2 K^2 \left( d\theta^2 + \sin ^2 \theta d\phi^2 \right) 
 + \frac{r_\infty ^2}{4K^2} \left( d\psi + \cos \theta d\phi \right)^2 , 
\label{met}
\\
 & A_\mu dx^\mu = \pm \frac{\sqrt{3 \rho_+ \rho_-} }{2 \rho} dt , 
\label{maxwellfield} 
\end{align}
with
\begin{align}
 F = \frac{\left( \rho - \rho_+ \right) \left( \rho - \rho_- \right) }{\rho^2 } , \qquad 
 K^2 = \frac{\rho + \rho_0}{\rho }  ,
\label{funcs} 
\end{align}
where the parameters $\rho_+ ,~ \rho_-$ and $\rho_0$ denote the outer and the inner horizons, and 
the typical scale of transition from five dimensions to effective four dimensions, respectively   
\cite{Matsuno:2009nz}. 
The parameter $r_\infty = 2 \sqrt{ \left( \rho _+ + \rho_0 \right) \left( \rho_- + \rho_0 \right) }$ 
gives the size of the compactified extra dimension $\psi$ at infinity. 
The coordinates run the ranges of 
$- \infty < t < \infty ,~ 0 < \rho < \infty ,~ 0 \leq \theta \leq \pi ,~ 0 \leq \phi \leq 2 \pi $ 
and $0 \leq \psi \leq 4 \pi $. 
The squashed Kaluza-Klein black hole solution is asymptotically locally flat, i.e., 
the metric asymptotes to a twisted constant S$^1$ fiber bundle over the four-dimensional Minkowski spacetime. 
In this paper, to avoid the existence of naked singularities on and outside the horizon, 
we restrict ourselves to the ranges of parameters such that
$\rho_+ \geq \rho_- \geq 0 ,~ \rho_- + \rho_0 > 0$.      
The Komar mass and the charge of the black hole are given by
\begin{align} 
 M = \frac{\pi r_\infty }{G_5} \left( \rho _+ + \rho_- \right) 
 = \frac{\rho _+ + \rho_- }{2 G_4} , 
 \qquad
 | Q | = \frac{2 \pi r_\infty }{G_5 } \sqrt{\rho _+ \rho_- } 
 = \frac{\sqrt{\rho _+ \rho_- }}{G_4 } , 
\end{align}
respectively, where the five-dimensional gravitational constant $G_5$ 
and the four-dimensional one $G_4$ are related as
$G_5 = 2 \pi r_\infty G_4$ \cite{Matsuno:2009nz}.  
In terms of the mass and the charge, the outer and the inner horizons are expressed as  
$\rho _\pm = G_4 \left( M \pm \sqrt{M^2 - Q^2} \right)$. 
The Hawking temperature and the entropy of the black hole 
associated with the surface gravity of the outer horizon $\kappa$ and 
the area of the outer horizon $\mathcal A$ are obtained as 
\begin{align}
 & T _\text{KK} = \frac{\kappa}{2 \pi}
 = \frac{\rho_+ - \rho_-}{4 \pi \rho_+ \sqrt{\rho_+ \left( \rho_+ + \rho_0 \right) } } , 
\label{htemperature0} 
\\  
 & S _\text{KK} = \frac{\mathcal A}{4 G_5} 
 = \frac{\pi \rho_+ \sqrt{\rho_+ \left( \rho_+ + \rho_0 \right) } }{G_4} , 
\end{align}
respectively, which satisfy the Smarr-type formula 
$M - 2 Q A_+ \left/ \sqrt{3} \right. = 2 T_\text{KK} S _\text{KK}$, 
where $| A_+ | = \left. \sqrt{3 \rho_- / \rho_+} \right/ 2$ 
is the gauge potential on the outer horizon 
\cite{Cai:2006td, Kurita:2007hu, Kurita:2008mj}.  
When $\rho _- = \rho _+$, the Hawking temperature $T_\text{KK}$ vanishes and 
the metric \eqref{met} represents the extremal charged Kaluza-Klein black hole 
with the mass $M = |Q|$ \cite{Ishihara:2005dp}.  
In the limit $\rho_0 \to 0$, we obtain the metric \eqref{met} with $K = 1$ 
which represents the four-dimensional Reissner-Nordstr\"{o}m black hole with a twisted constant S$^1$ fiber.  
We expect the appearance of the higher-dimensional corrections, which are related to the parameter $\rho_0$, 
to the tunneling radiation of particles across the black hole horizon and its Hawking temperature 
of four-dimensional relativity.

\section{Tunneling of scalar particles with quantum gravity effects}
\label{sec:stunneling}

\subsection{Derivation of modified Hawking temperature}

We consider the Hawking radiation by the tunneling of charged particles from 
the five-dimensional charged static squashed Kaluza-Klein black hole \eqref{met}, 
including the quantum gravity effects predicted by a generalized uncertainty principle. 
A number of generalized uncertainty principles are proposed 
as phenomenological models that would describe quantum gravity inspired 
uncertainty in the position of a particle in the vicinity of the black hole horizon 
\cite{Kempf:1994su, Kempf:1996nk, Isi:2013cxa}.
In this paper, we focus on the quadratic generalized uncertainty principle 
with the minimal measurable length in the form 
\cite{Kempf:1994su, Kempf:1996nk, Das:2008kaa} 
\begin{align}  
 \Delta x \Delta p \geq \frac{1}{2} \left[ 1 + \beta \left( \Delta p \right) ^2 \right] ,
\label{gup}
\end{align}  
where $\Delta x$ and $\Delta p$ respectively represent the uncertainties 
in the position and the momentum of a particle, 
$\beta = \beta_0 l_p^2 = \beta_0 \left/m_p^2 \right.$ is the parameter 
encoding quantum gravity effects on the particle dynamics,  
$\beta_0$ is a dimensionless positive deformation parameter,  
$l_p = \sqrt{\hbar G_4}$ is the Planck length and $m_p = \sqrt{\hbar / G_4}$ is the Planck mass.      
When $\Delta x \gg l_p$, one recovers the Heisenberg uncertainty principle as
$\Delta x \Delta p \geq 1 / 2$.   
In standard quantum mechanics, $\Delta x$ can be made arbitrarily small 
by letting $\Delta p$ grow correspondingly. 
However, this is no longer the case if the generalized uncertainty relation \eqref{gup} holds. 
When $\Delta p$ increases with decreasing $\Delta x$, 
the term $\beta \left( \Delta p \right) ^2$ on the right hand side of the uncertainty relation \eqref{gup} 
will eventually grow faster than the left hand side. 
Then $\Delta x$ can no longer be made arbitrarily small 
and there exists a minimal measurable length $\Delta x_0 = l_p \sqrt{\beta _0}$.   
The generalized uncertainty relation \eqref{gup} implies a small correction term 
to the commutation relation in the associative Heisenberg algebra for mirror-symmetric states, i.e., 
$\left[ x_i , p_j \right] = i \left( 1 + \beta g^{k l} p_k p_l \right) \delta_{i j}$, 
where $x_i = x_{0 i}$ is the position operator, 
$p_i = p_{0 i} \left( 1 + \beta g^{j k} p_{0 j} p_{0 k} \right)$ is the momentum operator,  
and 
$x_{0 i}$ and $p_{0 i}$ satisfy the canonical commutation relation  
$\left[ x_{0 i} , p_{0 j} \right] = i \delta_{i j}$ 
\cite{Kempf:1994su, Kempf:1996nk}.   
As the consequences of the generalized uncertainty principle \eqref{gup},  
using the energy mass shell condition 
$E^2 = g^{i j} p_i p_j + m ^2$, 
the modified energy operator takes the form  
$\tilde E = E \left( 1 - \beta m ^2 - \beta g^{i j} p_i p_j \right)$,  
where $E = i \left( \partial / \partial t \right)$ is the energy operator in standard quantum mechanics 
and $m$ is the mass of a particle 
\cite{Hossenfelder:2003jz}.

The Hawking radiation emitted by black holes may contain several kinds of particles.  
The choice of emitted particles is based on the prediction that 
the Hawking radiation would favor lower spin, lighter particles 
and it would be expected to be dominated by scalar particles 
\cite{Harris:2003eg, Sampaio:2009tp, Arbey:2021yke}. 
Then, in this section, we consider the tunneling of charged scalar particles 
across the horizon of the squashed Kaluza-Klein black hole with the quantum gravity effect.   
Using the momentum operator $p_i$ and the energy operator $\tilde E$, 
we obtain the modified Klein-Gordon equation with the Maxwell field 
up to the first order in the parameter $\beta$ \cite{Feng:2015jlj} as 
\begin{align}
& \left[ 
 g^{t \mu} \left( \hbar \partial _t + i e A_t \right) \left( \hbar \partial _\mu + i e A_\mu \right) 
 \right.
\notag \\
& \left. 
 + \left( g^{i j} \left( \hbar \partial _i + i e A_i \right) \left( \hbar \partial _j + i e A_j \right) - m^2 \right) 
 \left(1 - 2 \beta m^2 + 2 \beta \hbar ^2 g^{k l} \partial _k \partial _l \right) 
 \right] \Psi = 0 , 
\label{kgeq}
\end{align}
where $\partial _\mu = \partial / \partial x^\mu$, $e$ is the charge of the particle, 
$\Psi$ is the modified scalar field. 
When $A_\mu = 0$, the equation \eqref{kgeq} describes 
the modified Klein-Gordon equation without the Maxwell field \cite{Chen:2014xgj}.     
We assume that the wave function of the Klein-Gordon equation \eqref{kgeq} takes the form  
\begin{align}
 \Psi = \exp \left( \frac{i}{\hbar} I \left( x^\mu \right) \right) , 
\label{scalarfield} 
\end{align}
where $I$ is the action of the emitted scalar particle.  
Substituting the metric \eqref{met}, the Maxwell field \eqref{maxwellfield} and the ansatz \eqref{scalarfield} 
into the Klein-Gordon equation \eqref{kgeq}, 
the Wentzel-Kramers-Brillouin approximation to the leading order in $\hbar$ yields the equation of motion  
\begin{align}
 \frac{1}{F} \left( \partial _t I + e A_t \right) ^2 
 + \sigma \left( 2 \beta \sigma - 1 \right) = 0 , 
\label{kgeq2}
\end{align}
with
\begin{align}
 \sigma = m^2 + \frac{F}{K^2} \left( \partial _\rho I \right) ^2 
 + \frac{1}{\rho^2 K^2} \left( \partial _\theta I \right) ^2 
 + \frac{\left( \partial _\phi I - \cos \theta \partial _\psi I \right) ^2 }{\rho^2 K^2 \sin^2 \theta} 
 + \frac{4 K^2 }{r_\infty ^2} \left( \partial _\psi I \right) ^2 ,
\end{align}
where we relax the possible restriction on the parameter $\beta _0$ and 
regard the parameter $\beta$ as an independent variable.  
According to the Killing vector fields in the squashed Kaluza-Klein spacetime, 
$\partial / \partial t ,~ \partial / \partial \phi$ and $\partial / \partial \psi$, 
we consider the action $I$ in the form 
\begin{align}
  I = - \omega t + W (\rho , \theta) + J \phi +  L \psi , 
\label{action2} 
\end{align}
where $\omega ,~ J$ and $L$ are the emitted particle's energy and 
the angular momenta in the $\phi $ and the $\psi$ directions, respectively.  
Substituting the action \eqref{action2} into the equation \eqref{kgeq2},  
we obtain    
\begin{align}
 m^2 \left( 1 - 2 \beta m^2 \right) - \frac{1}{F} \left( \omega - e A_t \right) ^2 
 + \frac{\left( 1 - 4 \beta m^2 \right) F}{K^2} \left( \frac{\partial W}{\partial \rho} \right) ^2 
 - \frac{2 \beta F^2}{K^4} \left( \frac{\partial W}{\partial \rho} \right) ^4
 = 0 , 
\label{kgeq3}
\end{align}
where we restrict ourselves to the s-wave particles with 
$\theta =$ const, $J = 0 ,~ L = 0$,   
since the tunneling effect is a quantum one arising within the Planck length near the horizon region  
\cite{Parikh:1999mf, Srinivasan:1998ty, Angheben:2005rm, Kerner:2007rr, Kerner:2008qv}.     
From the equation \eqref{kgeq3}, we see that the function $W (\rho , \theta)$ can be written as 
$W (\rho , \theta) = R (\rho) + \Theta (\theta)$. 
Then, solving the equation \eqref{kgeq3} with the relation $\partial W / \partial \rho = d R/ d \rho$ 
on the black hole horizon $\rho = \rho_+$ yields the imaginary part of the action,  
\begin{align}
 \text{Im} R_\text{out} = - \text{Im} R_\text{in} = 
 \frac{\pi \rho_+ \sqrt{\rho_+ \left( \rho_+ + \rho_0 \right) } \left( \omega - e A_+ \right) }
 {\rho_+ - \rho_- } \left( 1 + \beta \Xi _s \right) + O \left( \beta^2 \right) , 
\end{align}
with 
\begin{align}
 \Xi _s =& ~ \frac{m^2}{2} 
 + \frac{\rho_+ \left( \omega - e A_+ \right) }{2 \left( \rho_+ + \rho_0 \right) \left( \rho_+ - \rho_- \right)^2 } 
 \left[ \omega \left( 4 \rho_+ \left( \rho_+ - 2 \rho_- \right) 
 + \rho_0 \left( 3 \rho_+ - 7 \rho_- \right) \right) 
 \right.
\notag \\
& + \left. e A_+ \left( 2 \rho_+ \left(\rho_+ + \rho_- \right) + \rho_0 \left( 3 \rho_+ + \rho_- \right) \right) 
 \right] ,
\label{scorrection} 
\end{align}
where the $\rho$-integral is performed by deforming the contour around the pole at the horizon  
which lies along the line of integration and gives $\pi i$ times the residue,  
and $R_\text{out}$ and $R_\text{in}$ correspond to the outgoing and the ingoing solutions, respectively. 
Then the tunneling probability amplitude of the charged scalar particles takes the form   
\begin{align}
 \Gamma 
 \simeq \frac{\exp \left( -2 \text{Im} R_\text{out} \right) }{\exp \left( -2 \text{Im} R_\text{in} \right) } 
 \simeq \exp \left( -\frac{4 \pi \rho_+ \sqrt{\rho_+ \left( \rho_+ + \rho_0 \right) } \left( 1 + \beta \Xi _s \right) }
 {\rho_+ - \rho_- } \left( \omega - e A_+ \right) \right) . 
\label{sprobamp} 
\end{align}
Thus, by comparing the probability amplitude \eqref{sprobamp} to the first order in the energy 
with the Boltzmann factor 
$\Gamma = \exp \left( - \left( \omega - e A_+ \right) \left/ T \right. \right)$   
in a thermal equilibrium state at the temperature $T$, 
we obtain the modified Hawking temperature of the squashed Kaluza-Klein black hole \eqref{met} as
\begin{align}
 T = T_\text{KK} \left( 1 - \beta \Xi _s \right) + O \left( \beta^2 \right) , 
\label{shtemperature} 
\end{align}
where the temperature $T_\text{KK}$ and the correction $\Xi _s$ are given by 
the equations \eqref{htemperature0} and \eqref{scorrection}, respectively.  
We see that the modified Hawking temperature \eqref{shtemperature} depends on 
the energy $\omega$, the mass $m$ and the charge $e$ of the emitted scalar particle, 
and is modified by the squashed Kaluza-Klein geometry, the Maxwell field and the generalized uncertainty principle 
through the parameters $\rho _0 ,~ \rho _-$ and $\beta$, respectively.

By taking some limits in the equation \eqref{shtemperature}, 
we obtain the Hawking temperatures of some five and four-dimensional black holes.  
First, when $\beta = 0$, the equation \eqref{shtemperature} coincides with 
the Hawking temperature of the five-dimensional squashed Kaluza-Klein black hole 
\cite{Cai:2006td, Kurita:2007hu, Kurita:2008mj}.  
Second, when $\rho _ - = 0$, 
introducing the new parameters 
$\rho _+ = r_+ ^2 \left/ \left( 2 \sqrt{r_\infty ^2 - r_+ ^2} \right) \right.$ and 
$\rho _0 = \left. \sqrt{r_\infty ^2 - r_+ ^2} \right/ 2$, 
then taking the limit $r_\infty \to \infty$, 
the equation \eqref{shtemperature} represents the modified Hawking temperature 
of the five-dimensional Schwarzschild-Tangherlini black hole obtained by the uncharged scalar particle tunneling 
\cite{Feng:2015jlj}. 
Lastly, when $\rho _0 = 0 ,~ \rho _ - = 0$, 
the equation \eqref{shtemperature} represents the modified Hawking temperature 
of the four-dimensional Schwarzschild black hole by the uncharged scalar particle tunneling  
\cite{Chen:2014xgj}.

\subsection{Evaporation of black holes}

\begin{figure}[!tbp]
\begin{center}
\includegraphics[scale=0.55]{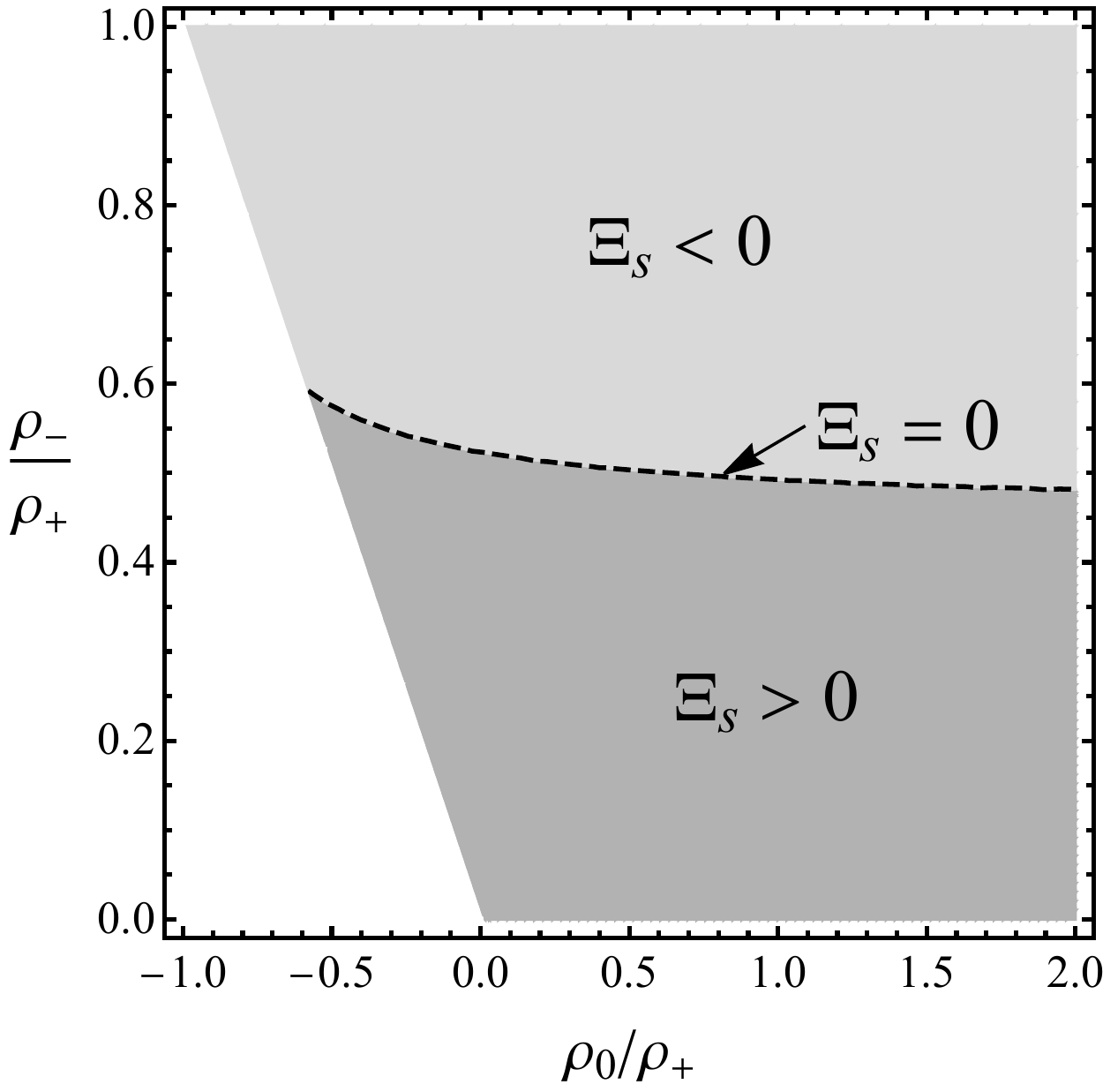} 
\caption{
Parameter regions of the correction \eqref{scorrection} with $\rho_+ \geq \rho_- \geq 0 ,~ \rho_- + \rho_0 > 0$ 
for $m=0 ,~ e / \omega = 0.1$.   
The positive $\Xi _s$ region (dark region) increases with increasing $e / \omega$.  
}
\label{fig:sregion}
\end{center}
\end{figure}

We consider the evaporation process of the squashed Kaluza-Klein black hole 
by the tunneling of scalar particles, 
which is affected by the correction $\Xi _s$ in the modified Hawking temperature $T$, 
as one of the quantum gravity effects in the Hawking radiation.     
Since all particles emitted by the Hawking radiation near the horizon region are effectively massless  
\cite{Parikh:1999mf, Srinivasan:1998ty, Angheben:2005rm, Kerner:2007rr, Kerner:2008qv}, 
the mass of the emitted scalar particle is not taken into account in the following discussion, i.e., $m = 0$. 
The mass of the squashed Kaluza-Klein black hole would decrease due to the radiation.   
When the black hole mass approaches the order of the Planck mass, 
the quantum gravity effect 
could be considered and we may discuss the value of the correction $\Xi _s$.  
First, when $\Xi _s = 0$, the effects related to the energy of the scalar particle, 
the asymptotically Kaluza-Klein structure and the Maxwell field are canceled. 
Then the Hawking temperature without the quantum gravity correction 
$T_\text{KK}$ appears and results in the complete evaporation.  
Second, when $\Xi _s < 0$, the temperature $T$ is higher than 
the temperature $T_\text{KK}$.  
Then the black hole accelerates the evaporation and there is no remnant left.  
Lastly, when $\Xi _s > 0$, the temperature $T$ is lower than 
the temperature $T_\text{KK}$.  
This implies that the combination of the effects related to the energy of the scalar particle, 
the compact extra dimension, the Maxwell field and the generalized uncertainty principle 
slows down the increase of the temperature due to the radiation.   
Then the evaporation may cease at the particular mass of the black hole and 
the black hole may be in a stable balanced state, leading to the remnant mass.  
We show the parameter regions of $\Xi _s$ in the figure \ref{fig:sregion}. 
From the equation \eqref{scorrection} and the figure \ref{fig:sregion}, 
we see that, for fixed $\rho _0 / \rho _+$ and $e / \omega$, 
the value of $\Xi _s$ decreases and changes from positive to negative as $\rho _- / \rho _+$ increases.  
Then the black holes with the large $Q$ would favor negative $\Xi _s$ and evaporate completely. 
We also find that the positive value of $\Xi _s$ and its parameter region decrease 
with increasing $\rho _0 / \rho _+$ for fixed $e / \omega$. 
Then the effectively four-dimensional black holes with the small $\rho _0$ and the small $Q$ 
would favor positive $\Xi _s$ and turn into the remnants at the final stage of the evaporation.

\begin{figure}[!tbp]
\begin{center}
\includegraphics[scale=0.58]{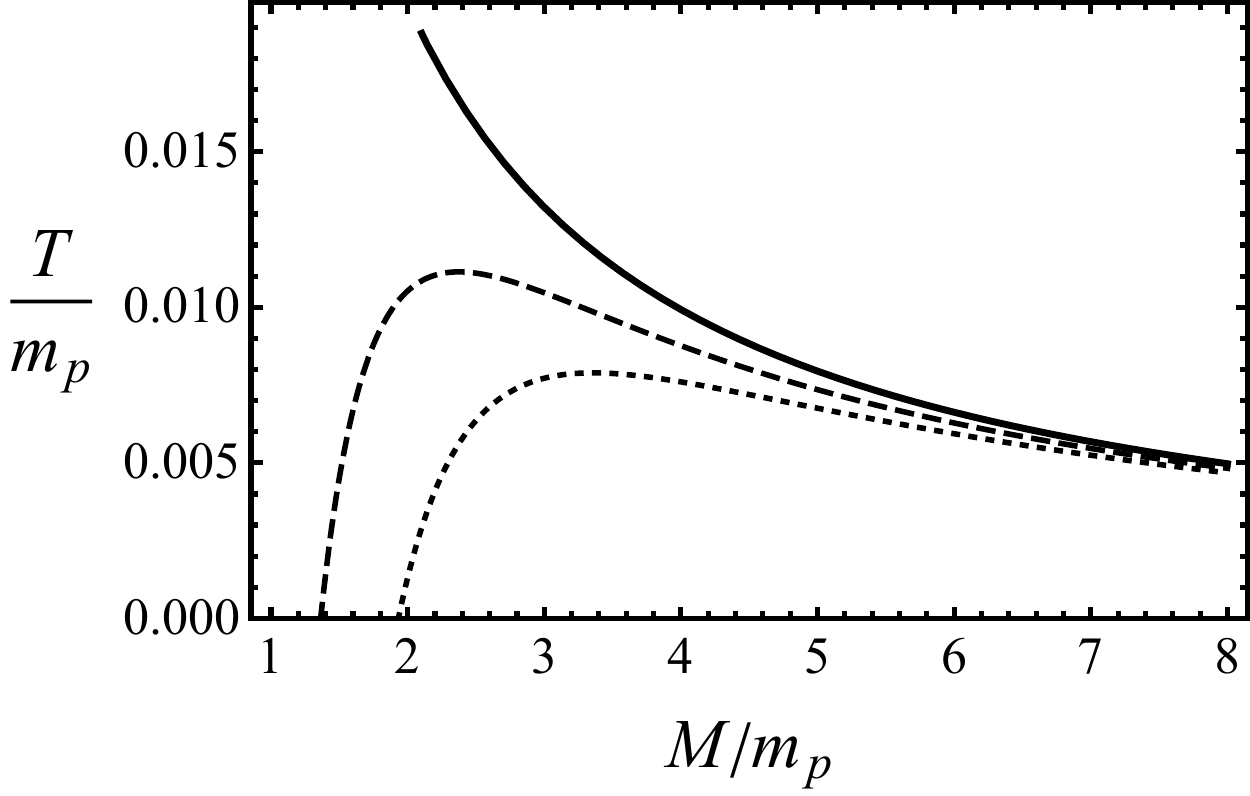} \qquad
\includegraphics[scale=0.58]{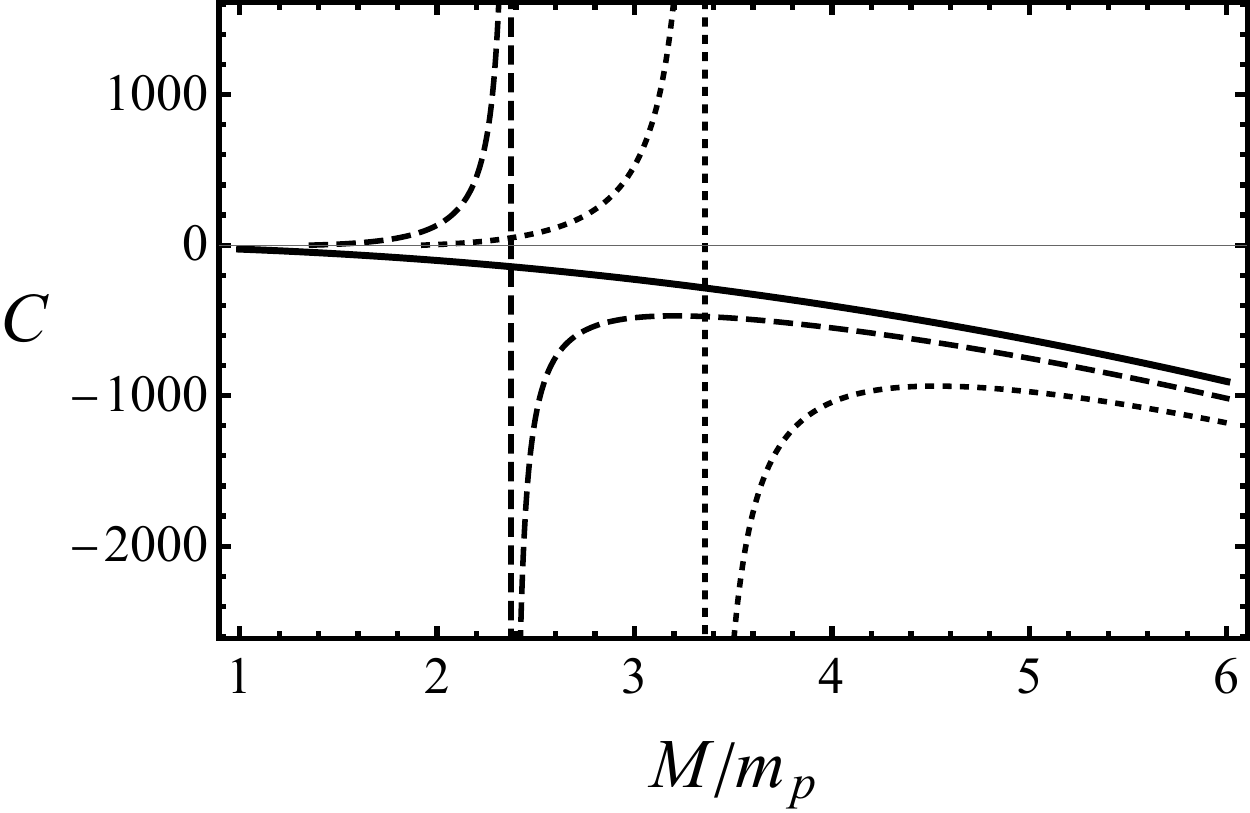} 
\caption{
Hawking temperatures \eqref{modtemp} and heat capacities \eqref{modcapa} 
in various $\beta_0$ for $r_\infty / l_p = 1$.   
$\beta_0 = 0$ (solid curves), $\beta_0 = 15$ 
($M_\text{rm} /m_p \simeq 1.36 ,~ M_\text{cr} /m_p \simeq 2.37$, dashed curves) 
and $\beta_0 = 30$ 
($M_\text{rm} /m_p \simeq 1.93 ,~ M_\text{cr} /m_p \simeq 3.36$, dotted curves). 
}
\label{fig:temperature}
\end{center}
\end{figure}

\begin{figure}[!tbp]
\begin{center}
\includegraphics[scale=0.58]{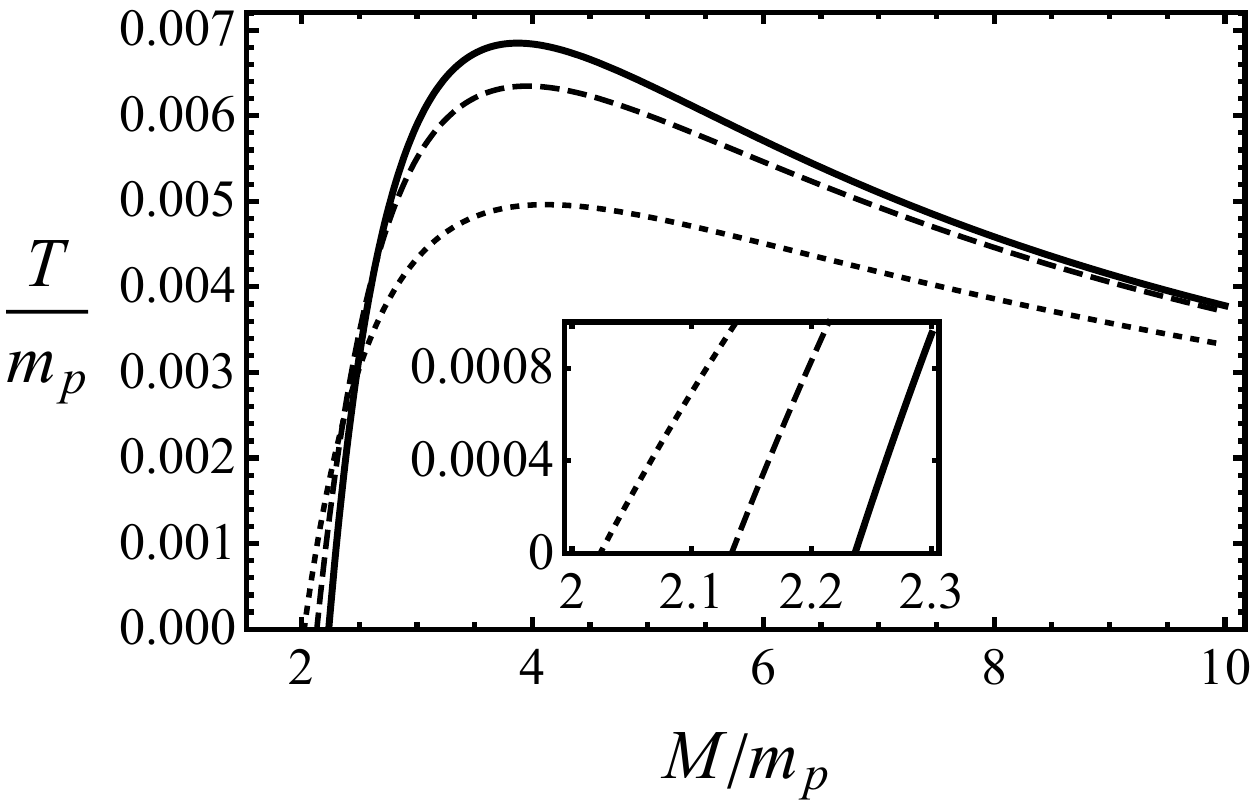} \qquad 
\includegraphics[scale=0.58]{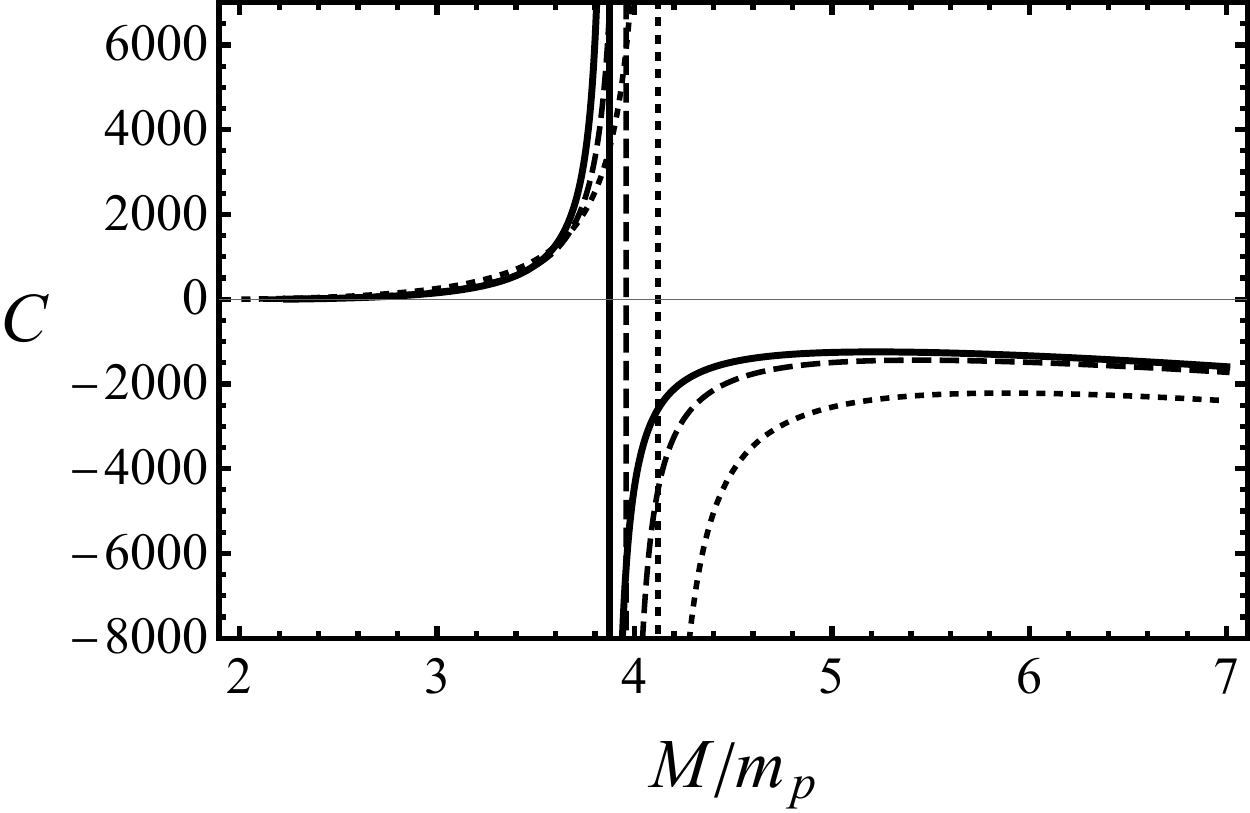}
\caption{
Hawking temperatures \eqref{modtemp} and heat capacities \eqref{modcapa}  
in various $r_\infty /l_p$ for $\beta_0 = 40$.   
$r_\infty = 0$ 
($M_\text{rm} /m_p \simeq 2.24 ,~ M_\text{cr} /m_p \simeq 3.87$, solid curves), 
$r_\infty / l_p = 8$ 
($M_\text{rm} /m_p \simeq 2.13 ,~ M_\text{cr} /m_p \simeq 3.96$, dashed curves) 
and $r_\infty / l_p = 25$ 
($M_\text{rm} /m_p \simeq 2.02 ,~ M_\text{cr} /m_p \simeq 4.12$, dotted curves). 
}
\label{fig:heatcapacity}
\end{center}
\end{figure}

Here, we restrict ourselves to the uncharged scalar particle radiation 
from the uncharged Kaluza-Klein black hole \eqref{met}, i.e., $\rho _- = 0$. 
In this case, the correction $\Xi _s$ is positive.  
Then, using the lower bound on the energy of the emitted particle 
$\omega \geq 1 / \Delta x$ \cite{Medved:2004yu} 
and the uncertainty in the position $x$ for the events near the black hole horizon $\Delta x \simeq 2 \rho_+$  
\cite{Medved:2004yu},  
we have the Hawking temperature \eqref{shtemperature} in terms of 
$M ,~ r_\infty ,~ \beta_0 ,~ m_p$ and $l_p$ as  
\begin{align}
 T = \frac{m_p }{4 \pi \sqrt{ \mu \left( 2 \mu + \sqrt{4 \mu^2 + \nu ^2} \right) } } 
 \left( 1 - \beta _0 \frac{10 \mu + 3 \sqrt{4 \mu^2 + \nu ^2} }
 {32 \mu^2 \left( 2 \mu + \sqrt{4 \mu^2 + \nu ^2} \right)} \right) 
 + O \left( \beta _0 ^2 \right) , 
\label{modtemp}
\end{align}
where $\mu = M / m_p$ and $\nu = r_\infty / l_p$. 
We find that, for the large black hole masses, 
since the quantum gravity effect is negligible at that scale, 
the modified temperature \eqref{modtemp} asymptotes to 
the temperature $T_\text{KK}$ with $\rho_- = 0$,  
which monotonically increases with decreasing $M / m_p$ for fixed $r_\infty / l_p$  
and is higher than the temperature \eqref{modtemp} with $\beta _0 \neq 0$.   
Using the temperature \eqref{modtemp} and the Smarr-type formula, 
we have the entropy with the quantum gravity effect as  
\begin{align}
 S = 2 \pi \mu \sqrt{ \mu \left( 2 \mu + \sqrt{4 \mu^2 + \nu ^2} \right) } 
 \left( 1 + \beta_0 \frac{10 \mu + 3 \sqrt{4 \mu^2 + \nu ^2} }
 {32 \mu ^2 \left( 2 \mu + \sqrt{4 \mu^2 + \nu ^2} \right) } \right)
 + O \left( \beta _0 ^2 \right) . 
\label{modentropy}
\end{align}
When the deformation parameter of the generalized uncertainty principle vanishes, i.e., $\beta _0 = 0$, 
the equation \eqref{modentropy} coincides with 
the entropy $S_\text{KK}$
with $\rho _- =0$.  
Using the temperature \eqref{modtemp} and the entropy \eqref{modentropy},  
we obtain the modified heat capacity $C = T \left( \partial S / \partial T \right)$ as  
\begin{align}
 C = & ~ \pi  \left(2048 \mu ^5 +448 \mu ^3 \nu ^2 -2 \beta _0 \mu \nu ^2 
   + \left( 1024 \mu ^4 +96 \mu ^2 \nu ^2 -3 \beta _0 \nu ^2 \right) \sqrt{4 \mu ^2+\nu ^2} \right) 
\notag \\
   & \times \left( 64 \mu ^3 -10 \beta _0 \mu + \left( 32 \mu ^2 - 3 \beta _0 \right) \sqrt{4 \mu ^2+\nu ^2}
   \right) \sqrt{\mu \left(2 \mu + \sqrt{4 \mu ^2+\nu ^2} \right)} 
\notag \\   
   & \times \left[ 48 \beta _0 \mu  \left(512 \mu ^4+136 \mu ^2 \nu ^2 +5 \nu ^4 
   + \left(256 \mu ^3 +36 \mu \nu ^2 \right) \sqrt{4 \mu ^2+\nu ^2} \right) \right. 
\notag \\
   & \left. -512 \mu ^3 \left(128 \mu ^4+32 \mu ^2 \nu ^2 +\nu ^4 
   + \left( 64 \mu ^3 + 8 \mu \nu ^2 \right) \sqrt{4 \mu ^2+\nu ^2} \right) \right] ^{-1} ,
\label{modcapa}
\end{align}
which gives the information on the thermodynamic stability of the present system.  
We find that 
the heat capacity \eqref{modcapa} with $\beta_0 = 0$ 
monotonically increases with decreasing $M / m_p$ for fixed $r_\infty / l_p$ 
and is negative for $M > 0$.  
We show the behaviors of the Hawking temperature $T / m_p$ and the heat capacity $C$ 
versus $M / m_p$ in the figures \ref{fig:temperature} and \ref{fig:heatcapacity}.   
From the left panels of the figures \ref{fig:temperature} and \ref{fig:heatcapacity},  
we see that, as the black hole mass decreases,  
the modified temperature \eqref{modtemp} with $\beta _0 \neq 0$ reaches 
the local maximum value at the critical mass  
\begin{align}
 M_\text{cr} = \frac{m_p}{4} \sqrt{
 \frac{ 6 \beta_0 \left( 3 \beta_0 - 5 \nu ^2 \right) + 3 \beta_0 \sqrt{6 \beta_0 \left(6 \beta_0 + 5 \nu ^2 \right)} }
 {2 \left( 3 \beta_0 - 2 \nu ^2 \right)} } , 
\label{crmass}
\end{align}
and then decreases to zero at the minimum value of the mass 
\begin{align}
 M_\text{rm} = \frac{m_p}{4} \sqrt{
 \frac{ 2 \beta_0 \left( \beta_0 + 3 \nu ^2 \right) + \beta_0 \sqrt{2 \beta_0 \left(2 \beta_0 + 3 \nu ^2 \right)} }
 {2 \left( \beta_0 + 2 \nu ^2 \right)} } . 
\label{remmass}
\end{align}
The existence of local maximum and vanishing values of the Hawking temperature are related to 
a phase transition of the black hole and an evaporation remnant.   
From the right panels of the figures \ref{fig:temperature} and \ref{fig:heatcapacity},
we find that there exist three regimes in the heat capacity \eqref{modcapa} with $\beta_0 \neq 0$, 
i.e., $C < 0$ for $M > M_\text{cr}$, $C \geq 0$ for $M_\text{rm} \leq M < M_\text{cr}$,   
and $C \to \infty$ for $M \to M_\text{cr}$.   
The discontinuity between negative and positive values of the heat capacity 
indicates a phase transition from a thermodynamic unstable phase to a stable one.   
Then we see that, 
at the local maximum temperature specified by the mass $M_\text{cr}$,  
the system undergoes a transition from an unstable negative heat capacity phase 
to a stable positive heat capacity cooling down towards a cold extremal configuration 
with the mass $M_\text{rm}$ due to the radiation.  
At the minimum mass $M_\text{rm}$, 
the black hole no longer radiates since the Hawking temperature \eqref{modtemp} vanishes.    
At the same time, the heat capacity \eqref{modcapa} also vanishes,   
which implies that the black hole may not exchange its energy with the surrounding environment.   
Then the generalized uncertainty principle prevents the squashed Kaluza-Klein black hole 
to completely evaporate and results in the thermodynamic stable remnant, 
similar to the evaporation of the black holes in the noncommutative model and the asymptotically safe gravity 
\cite{Bonanno:2000ep, Myung:2006mz, DiGennaro:2021vev}.   
From the left panels of the figures \ref{fig:temperature} and \ref{fig:heatcapacity}, 
we see that the mass of the black hole remnant $M_\text{rm}$ 
increases with increasing $\beta_0$ for fixed $r_\infty / l_p$, 
while it decreases with increasing $r_\infty / l_p$ for fixed $\beta_0$.   
When $r_\infty = 0$, the equation \eqref{remmass} represents 
the mass of the four-dimensional Schwarzschild black hole remnant 
\cite{Feng:2015jlj}.  
If the deformation parameter is $\beta _0 \simeq 1$ 
\cite{Feng:2015jlj} 
and the size of the extra dimension is $r_\infty \simeq 0.1$ mm 
\cite{Matsuno:2009nz}, 
we can estimate that the mass of the squashed Kaluza-Klein black hole remnant is 
$M _\text{rm} \simeq 10 ^{-8}$ kg, which is of order the Planck mass 
like the black hole remnants in the minimally geometric deformation model, the quadratic gravity 
and the asymptotically safe gravity 
\cite{Casadio:2017sze, Kuntz:2019gka, Bonanno:2000ep}.

\begin{figure}[!tbp]
\begin{center}
\includegraphics[scale=0.55]{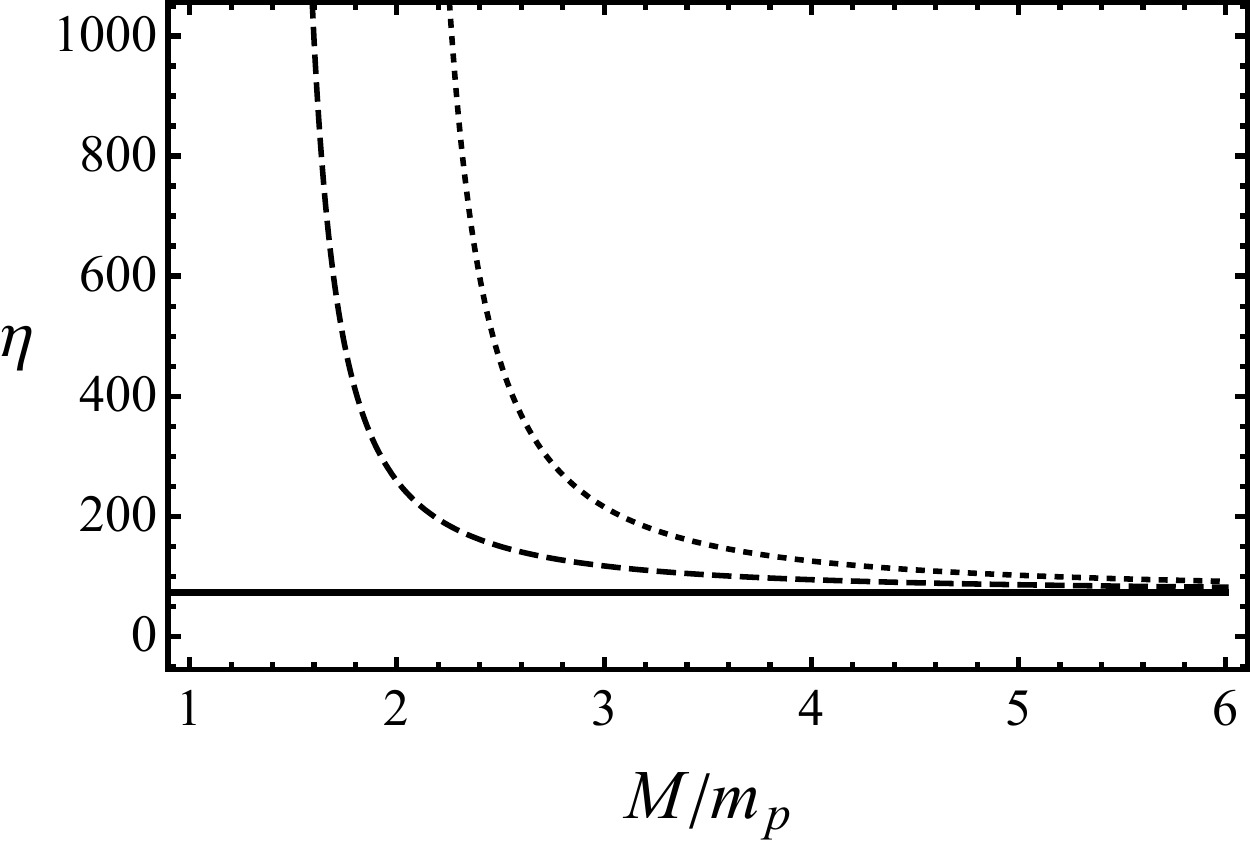} \qquad
\includegraphics[scale=0.55]{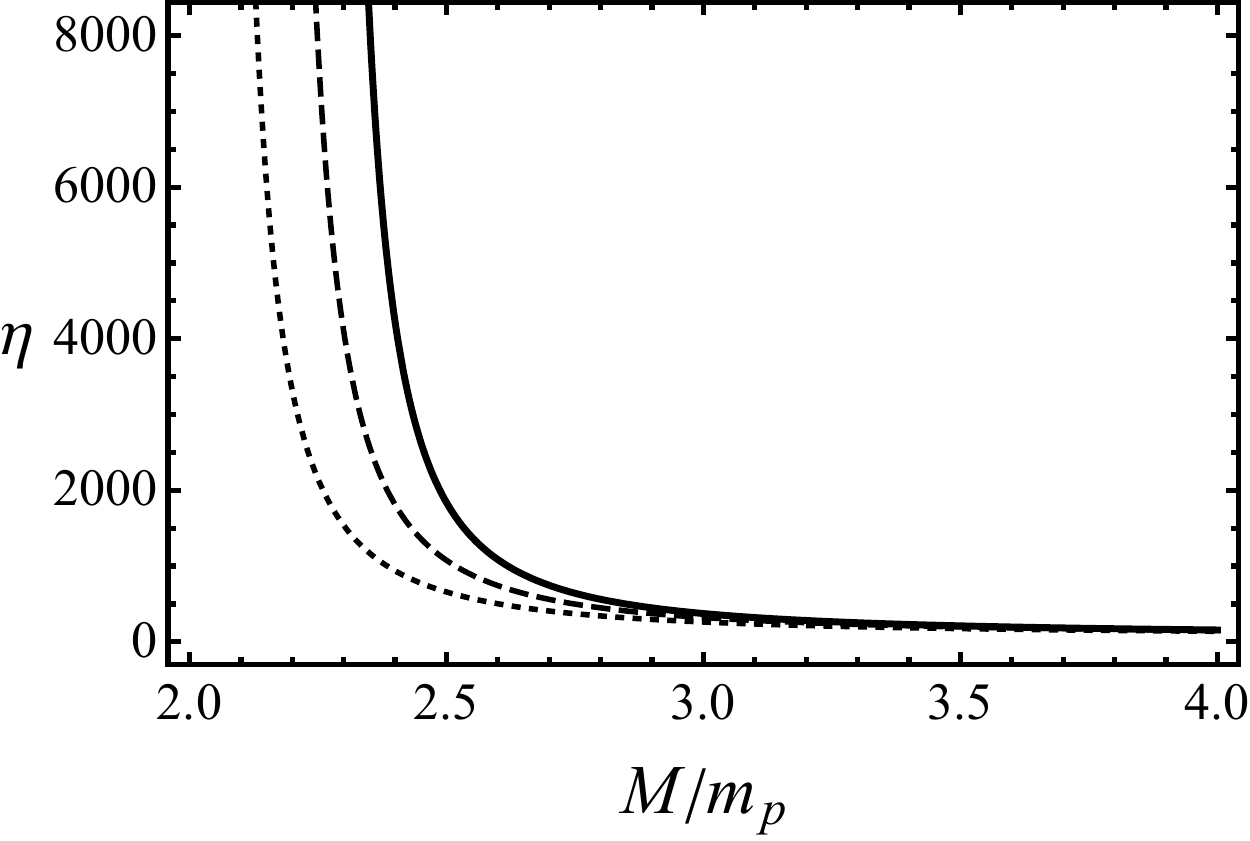} 
\caption{
Sparsities of Hawking radiation \eqref{modspar} in various $\beta_0$ for $r_\infty / l_p = 1$ (left panel).   
$\beta_0 = 0$ (solid line), 
$\beta_0 = 15$ ($M_\text{rm} /m_p \simeq 1.36$, dashed curve) 
and $\beta_0 = 30$ ($M_\text{rm} /m_p \simeq 1.93$, dotted curve). 
The same ones in various $r_\infty /l_p$ for $\beta_0 = 40$ (right panel).   
$r_\infty = 0$ ($M_\text{rm} /m_p \simeq 2.24$, solid curve), 
$r_\infty / l_p = 8$ ($M_\text{rm} /m_p \simeq 2.13$, dashed curve) 
and $r_\infty / l_p = 25$ ($M_\text{rm} /m_p \simeq 2.02$, dotted curve). 
}
\label{fig:sparsity}
\end{center}
\end{figure}

We consider the impact of the quantum gravity effect 
onto the sparsity of the Hawking radiation during the evaporation process 
of the squashed Kaluza-Klein black hole. 
Sparsity is defined by the average time gap between two successive emissions of the particle  
over the characteristic timescale of the emission of the individual particle 
\cite{Page:1976df, Gray:2015pma}.    
It has been shown that the Hawking radiation is sparse 
throughout the whole Hawking evaporation process of the four-dimensional black hole 
\cite{Gray:2015pma}.  
However, it has also been shown that the sparsity of the Hawking radiation would be modified by 
some quantum gravity effects in some higher-dimensional black hole spacetimes 
\cite{Alonso-Serrano:2018ycq, Ong:2018syk, Alonso-Serrano:2020hpb, DiGennaro:2021vev, 
Hod:2016rmg, Paul:2016xvb, Miao:2017jtr, Feng:2018jqf, Schuster:2019xvp}.  
Further, once the wave effect of the radiation is taken into account, 
the sparsity becomes a crucial feature that extends the black hole lifetime.  
Then we consider the sparsity of the Hawking radiation in the form  
$\eta = \left. \lambda ^2 \right/ \mathcal A _\text{eff}$, 
where $\lambda = 2 \pi \left/ T \right.$ is the thermal wavelength of the Hawking particle, 
$\mathcal A _\text{eff} = \left. 27 \tilde{\mathcal A} \right/ 4$ is the effective area 
that corresponds to the universal cross section at high frequencies 
and $\tilde{\mathcal A}$ is the area of the black hole horizon 
\cite{Alonso-Serrano:2018ycq, Ong:2018syk, Alonso-Serrano:2020hpb, Feng:2018jqf}.   
If the sparsity is much less than $1$, 
the Hawking radiation is a typical blackbody radiation 
where its thermal wavelength is much shorter than the size of the emitting body.  
On the other hand, if the sparsity is much greater than $1$, 
the Hawking radiation is not a continuous emission of particles but a sparse radiation, i.e.,  
most particles are randomly emitted in a discrete manner with pauses in between.  
Using the temperature \eqref{modtemp} and 
the area of the horizon $\tilde{\mathcal A} = 4 \pi \mu l_p ^2 \left( 2 \mu + \sqrt{4 \mu ^2+\nu ^2} \right)$, 
we obtain the modified sparsity of the Hawking radiation as  
\begin{align}
 \eta = \frac{65536 \pi ^3 \mu ^4 \left( 2 \mu + \sqrt{4 \mu ^2+\nu ^2} \right) ^2}
 {27 \left( 64 \mu ^3 -10 \beta _0 \mu + \left( 32 \mu ^2 - 3 \beta _0 \right) \sqrt{4 \mu ^2+\nu ^2} \right) ^2 }.
\label{modspar}
\end{align}
We see that the sparsity \eqref{modspar} depends on the black hole mass $M$, 
the extra dimension size $r _\infty$ and the deformation parameter $\beta _0$.  
When $\beta _0 = 0$, the equation \eqref{modspar} coincides with 
the sparsity of the Hawking radiation from the four-dimensional Schwarzschild black hole, 
which keeps a constant value $\left. 64 \pi ^3 \right/ 27$ during the whole evaporation process  
\cite{Alonso-Serrano:2018ycq}.   
We show the behaviors of the sparsity $\eta$ versus $M / m_p$ in the figure \ref{fig:sparsity}. 
From the figure \ref{fig:sparsity}, we see that 
the sparsity \eqref{modspar} with $\beta _0 \neq 0$ increases with decreasing $M / m_p$ 
for fixed $r_\infty / l_p$ and $\beta _0$, and then diverges at $M = M_\text{rm}$. 
This implies that the average time between the emission of the successive Hawking quanta 
becomes much larger than the timescales set by the energies of the emitted quanta, 
so the radiation is very sparse.  
Then the sparsity of the Hawking radiation is enhanced due to the quantum gravity effect.  
Thus, 
similar to the evaporation of the black holes in the noncommutative model and the asymptotically safe gravity 
\cite{Bonanno:2000ep, DiGennaro:2021vev},  
the squashed Kaluza-Klein black hole with the generalized uncertainty principle 
would take an infinite amount of time to radiate a particle at the final stage of the evaporation, 
and then turn into the remnant when the black hole mass $M$ approaches the mass $M_\text{rm}$.  
Since the squashed Kaluza-Klein black hole remnant would not radiate 
and its gravitational interaction would be very weak, 
it would be difficult to observe the remnants in our Universe directly. 
However, it would be expected that one possible indirect signature of the black hole remnant 
might be associated with the cosmic gravitational wave background 
\cite{Chen:2002tu, Chen:2004ft}.

\section{Tunneling of fermions with quantum gravity effects}
\label{sec:ftunneling}

We consider the Hawking radiation by the tunneling of charged fermions 
in the five-dimensional charged static squashed Kaluza-Klein black hole spacetime \eqref{met} 
with the quantum gravity effects.    
As the consequences of the generalized uncertainty principle \eqref{gup},  
using the momentum operator $p_i$ and the energy operator $\tilde E$, 
which leads effectively to a replacement of third order time derivatives by spatial derivatives,   
we obtain the modified Dirac equation with the Maxwell field up to the first order in the parameter $\beta$ 
\cite{Hossenfelder:2003jz, Chen:2013ssa} 
which describes the behaviors of spinning particles in curved spacetimes as   
\begin{align}  
& \left[ 
 i \hbar \gamma ^t \partial _t 
 + \left( i \hbar \left( \gamma ^i \partial _i + \gamma^\mu \Omega_\mu \right) - e \gamma^\mu A_\mu + m \right) 
 \left( 1 - \beta m^2 + \beta \hbar ^2 g^{j k} \partial _j \partial _k \right) 
\right] \tilde \Psi = 0 , 
\label{diraceq}
\end{align}  
where 
$\tilde \Psi$ is the modified Dirac field, 
$\Omega _\mu = i \left. \omega_\mu ^{a b} \Sigma _{a b} \right/ 2$, 
$\omega_{\mu ~b} ^{~a} = e_\nu ^a e_b ^\lambda \Gamma _{\mu \lambda} ^\nu 
- e_b ^\nu \partial_\mu e_\nu ^a$ is the spin connection defined by 
the Christoffel symbol $\Gamma _{\mu \nu } ^\lambda $ and the vielbein $e_\mu ^a$, and 
$\Sigma ^{a b} = i \left. \left[ \gamma ^a , \gamma ^b \right] \right/ 4$ is the Lorentz spinor generator.  
The vielbeins $e_\mu ^a$ and the gamma matrices $\gamma ^\mu = e_a ^\mu \gamma^a$ 
satisfy the conditions $g^{\mu \nu} e_\mu ^a e_\nu ^b = \eta ^{a b} = \text{diag} \left( -1 , 1 , 1 , 1 , 1 \right)$ 
and $\left \{ \gamma ^\mu , \gamma ^\nu \right \} = 2 g^{\mu \nu}$, respectively.    
When $A _\mu = 0$, the equation \eqref{diraceq} describes 
the modified Dirac equation without the Maxwell field  
\cite{Chen:2013pra}.

For a spin-$1/2$ fermion, there are a spin-up and a spin-down states.  
In this paper, we only consider the tunneling radiation of the fermion with the spin-up state,  
since the discussion of the spin-down state is fully analogous.   
For the spin-up state, 
we employ the following ansatz for the spinor field describing the fermion, 
\begin{align}
 \tilde \Psi = 
\begin{pmatrix}
 U (x^\mu) \\ 0 \\ V (x^\mu) \\ 0 \\
\end{pmatrix} 
 \exp \left( \frac{i}{\hbar} \tilde I (x^\mu) \right) , 
\label{spinorfield}
\end{align}
with the action $\tilde I$ and the functions $U$ and $V$.   
The corresponding gamma matrices of the squashed Kaluza-Klein spacetime \eqref{met} are  
\begin{align}
& \gamma^t = \frac{1}{\sqrt{F} }
\begin{pmatrix}
 0 & 0 & 1 & 0 \\
 0 & 0 & 0 & 1 \\
 -1 & 0 & 0 & 0 \\
 0 & -1 & 0 & 0 \\
\end{pmatrix} ,
\qquad
\gamma^{\rho} = \frac{\sqrt{F} }{K}
\begin{pmatrix}
 0 & 0 & 1 & 0 \\
 0 & 0 & 0 & -1 \\
 1 & 0 & 0 & 0 \\
 0 & -1 & 0 & 0 \\
\end{pmatrix} ,
\qquad
\gamma^{\theta} = \frac{1}{\rho K}
\begin{pmatrix}
 0 & 0 & 0 & 1 \\
 0 & 0 & 1 & 0 \\
 0 & 1 & 0 & 0 \\
 1 & 0 & 0 & 0 \\
\end{pmatrix} ,
\notag \\
& \gamma^{\phi} = \frac{1}{\rho K \sin \theta }
\begin{pmatrix}
 0 & 0 & 0 & -i \\
 0 & 0 & i & 0 \\
 0 & -i & 0 & 0 \\
 i & 0 & 0 & 0 \\
\end{pmatrix} ,
\qquad
\gamma^{\psi} = 
\begin{pmatrix}
 -2 K / r_\infty & 0 & 0 & i \tilde K \\
 0 & -2 K / r_\infty & -i \tilde K & 0 \\
 0 & i \tilde K & 2 K / r_\infty & 0 \\
 -i \tilde K & 0 & 0 & 2 K / r_\infty \\
\end{pmatrix} ,
\label{gammamatrices}
\end{align}
where $\tilde K = \cos \theta \left/ \left(\rho K \sin \theta \right) \right.$  
\cite{Li:2011zzm, Stetsko:2014dda}.  
Substituting the metric \eqref{met}, the Maxwell field \eqref{maxwellfield},  
the ansatz \eqref{spinorfield} and the matrices \eqref{gammamatrices} 
into the Dirac equation \eqref{diraceq}, 
the Wentzel-Kramers-Brillouin approximation to the leading order in $\hbar$ yields the equations of motion  
\begin{align}
& U \left( \frac{1}{\sqrt{F} } \left( \partial _t \tilde I + e A_t \tilde \sigma \right) 
 - \tilde \sigma \frac{\sqrt{F} }{K} \partial _\rho \tilde I \right)
 - \tilde \sigma V \left( \frac{2 K}{r_\infty} \partial _\psi \tilde I - m \right) = 0 ,
\label{eom1} 
 \\
& V \left( \frac{1}{\sqrt{F} } \left( \partial _t \tilde I + e A_t \tilde \sigma \right) 
 + \tilde \sigma \frac{\sqrt{F} }{K} \partial _\rho \tilde I \right) 
 - \tilde \sigma U \left( \frac{2 K}{r_\infty} \partial _\psi \tilde I + m \right) = 0 , 
\label{eom2}  
 \\
& \tilde \sigma \left( \partial _\theta \tilde I 
 + \frac{i}{\sin \theta} \left( \partial _\phi \tilde I - \cos \theta \partial _\psi \tilde I \right) \right) = 0 , 
\label{eom3}  
\end{align}
with 
\begin{align}
 \tilde \sigma = 1 - \beta m^2 - \frac{\beta F}{K^2} \left( \partial _\rho \tilde I \right) ^2 
 - \frac{\beta }{\rho^2 K^2} \left( \partial _\theta \tilde I \right) ^2 
 - \frac{\beta \left( \partial _\phi \tilde I - \cos \theta \partial _\psi \tilde I \right) ^2 }{\rho^2 K^2 \sin^2 \theta} 
 - \frac{4 \beta K^2 }{r_\infty ^2} \left( \partial _\psi \tilde I \right) ^2 ,
\end{align}
where we relax the possible restriction on the parameter $\beta _0$ and 
regard the parameter $\beta$ as an independent variable.  
It would be difficult to solve the equations \eqref{eom1}-\eqref{eom3} directly. 
Then, according to the Killing vector fields in the squashed Kaluza-Klein spacetime, 
we consider the action $\tilde I$ in the form 
\begin{align}
 \tilde I = - \tilde \omega t + \tilde W (\rho , \theta) + \tilde J \phi + \tilde L \psi , 
\label{action} 
\end{align}
where $\tilde \omega ,~ \tilde J$ and $\tilde L$ are the emitted fermion's energy and 
the angular momenta in the $\phi $ and the $\psi$ directions, respectively. 
Substituting the action \eqref{action} into the equation \eqref{eom3}, we have 
\begin{align}
 & \left( \frac{\partial \tilde W}{\partial \theta}
 + \frac{i}{\sin \theta} \left( \tilde J - \tilde L \cos \theta \right) \right)
\notag \\
 & \times \left[ 1 - \beta m^2 - \frac{\beta F}{K^2} \left( \frac{\partial \tilde W}{\partial \rho} \right) ^2 
 - \frac{\beta }{\rho^2 K^2} \left( \frac{\partial \tilde W}{\partial \theta} \right) ^2 
 - \frac{\beta \left( \tilde J - \tilde L \cos \theta \right) ^2 }{\rho^2 K^2 \sin^2 \theta} 
 - \frac{4 \beta K^2 \tilde L^2}{r_\infty ^2} \right] 
 = 0 .
\label{eom30}  
\end{align}
Since the parameter $\beta$ is a small quantity which represents the quantum gravity effects 
by the generalized uncertainty principle, 
the expression inside the square brackets in the equation \eqref{eom30} cannot vanish. 
Then we obtain  
\begin{align}
 \frac{\partial \tilde W}{\partial \theta}
 + \frac{i}{\sin \theta} \left( \tilde J - \tilde L \cos \theta \right) = 0 .
\label{eom31}  
\end{align}
Substituting the action \eqref{action} and the equation \eqref{eom31} 
into the equations \eqref{eom1} and \eqref{eom2}, 
and cancelling the functions $U$ and $V$, we have    
\begin{align}
& \left( \frac{4 K^2 \tilde L^2 }{r_\infty ^2} - \frac{e^2 A_t ^2 }{F} -m^2 \right) 
   \left( \frac{4 \beta K^2 \tilde L^2}{r_\infty ^2} + \beta  m^2 -1 \right)^2 
   - \frac{2 \tilde \omega e A_t }{F}  \left( \frac{4 \beta K^2 \tilde L^2 }{r_\infty ^2}+\beta  m^2-1\right) 
   - \frac{\tilde \omega ^2}{F} 
\notag \\   
& + \left( \frac{F}{K^2} \left( \frac{4 \beta K^2 \tilde L^2 }{r_\infty ^2} +\beta m^2 -1 \right) 
   \left( \frac{12 \beta K^2 \tilde L^2}{r_\infty ^2} -\frac{2 \beta e^2 A_t ^2}{F} -\beta m^2 -1 \right)
   - \frac{2 \tilde \omega \beta e A_t }{K^2} \right) \left( \frac{\partial \tilde W}{\partial \rho} \right)^2
\notag \\
& + \frac{\beta F^2}{K^4} \left( \frac{12 \beta K^2 \tilde L^2 }{r_\infty ^2} 
   -\frac{\beta e^2 A_t ^2 }{F} + \beta m^2 -2 \right) \left( \frac{\partial \tilde W}{\partial \rho} \right)^4 
   + \frac{\beta ^2 F^3}{K^6} \left( \frac{\partial \tilde W}{\partial \rho} \right)^6
 = 0 .
\label{eom4} 
\end{align}
From the equations \eqref{eom31} and \eqref{eom4}, 
we see that the function $\tilde W (\rho , \theta)$ can be written as 
$\tilde W (\rho , \theta) = \tilde R (\rho) + \tilde \Theta (\theta)$. 
Then, from the equation \eqref{eom31} with the relation 
$\partial \tilde W / \partial \theta = d \tilde \Theta / d \theta$, 
we find that $\tilde \Theta$ must be a complex function.  
Neglecting the higher-order terms of $\beta$ in the equation \eqref{eom4} 
with the relation $\partial \tilde W / \partial \rho = d \tilde R / d \rho$
and solving the obtained equation on the black hole horizon $\rho = \rho_+$ 
for the classically forbidden trajectory yield the imaginary part of the action,  
\begin{align}
 \text{Im} \tilde R_\text{out} = - \text{Im} \tilde R_\text{in} = 
 \frac{\pi \rho_+ \sqrt{\rho_+ \left( \rho_+ + \rho_0 \right) } \left( \tilde \omega - e A_+ \right) }
 {\rho_+ - \rho_- } \left( 1 + \beta \Xi _f \right) + O \left( \beta^2 \right) , 
\end{align}
with 
\begin{align}
\Xi _f = & ~ \frac{3 \tilde \omega m^2 }{2 \left( \tilde \omega - e A_+ \right) } 
 + \frac{\tilde \omega \tilde L^2 }{2 \rho_+ \left( \rho_- + \rho_0 \right) \left( \tilde \omega - e A_+ \right) } 
\notag \\
& + \frac{\tilde \omega ^2 \rho_+ \left( 4 \rho_+ \left( \rho_+ - 2 \rho_- \right) 
 + \rho_0 \left( 3 \rho_+ - 7 \rho_- \right) \right) 
 + e \tilde \omega \rho_+ A_+ \left( 4 \rho_+ \rho_- + \rho_0 \left( \rho_+ + 3 \rho_- \right) \right) }
 {2 \left( \rho_+ + \rho_0 \right) \left( \rho_+ - \rho_- \right)^2 }  , 
\label{fcorrection} 
\end{align}
where $\tilde R_\text{out}$ and $\tilde R_\text{in}$ correspond to the outgoing and the ingoing solutions, respectively. 
Then the tunneling probability amplitude of the charged fermions reads 
\begin{align}
 \tilde \Gamma 
 \simeq \frac{\exp \left( -2 \text{Im} \tilde R_\text{out} \right) }{\exp \left( -2 \text{Im} \tilde R_\text{in} \right) } 
 \simeq \exp \left( -\frac{4 \pi \rho_+ \sqrt{\rho_+ \left( \rho_+ + \rho_0 \right) } \left( 1 + \beta \Xi _f \right) }
 {\rho_+ - \rho_- } \left( \tilde \omega - e A_+ \right) \right) , 
\label{probamp} 
\end{align}
where the contribution from the function $\tilde \Theta$ cancels out upon dividing the outgoing probability 
by the ingoing one, since the same solution for $\tilde \Theta$ is obtained for 
both the outgoing and the ingoing cases. 
Thus, by comparing the probability amplitude \eqref{probamp} to the first order in the energy 
with the Boltzmann factor 
$\tilde \Gamma = \exp \left( \left. - \left( \tilde \omega - e A_+ \right) \right/ \tilde T \right)$   
in a thermal equilibrium state at the temperature $\tilde T$, 
we obtain the modified Hawking temperature of the squashed Kaluza-Klein black hole \eqref{met} as
\begin{align}
 \tilde T = T_\text{KK} \left( 1 - \beta \Xi _f \right) + O \left( \beta^2 \right) , 
\label{htemperature} 
\end{align}
where the temperature $T_\text{KK}$ and the correction $\Xi _f$ are given by 
the equations \eqref{htemperature0} and \eqref{fcorrection}, respectively.  
We see that the modified Hawking temperature \eqref{htemperature} is determined by 
the mass $M$ and the charge $Q$ of the squashed Kaluza-Klein black hole, 
the size of the extra dimension $r_\infty$, the deformation parameter $\beta _0$,  
and the energy $\tilde \omega$, the mass $m$, the charge $e$ and the angular momentum $\tilde L$ 
of the emitted fermion.

By taking some limits in the equation \eqref{htemperature}, 
we obtain the Hawking temperatures of some five and four-dimensional black holes.  
First, when $\beta = 0$, the equation \eqref{htemperature} coincides with 
the Hawking temperature of the five-dimensional squashed Kaluza-Klein black hole 
\cite{Cai:2006td, Kurita:2007hu, Kurita:2008mj}.  
Second, when $\rho _ - = 0 ,~ \tilde L = 0$, 
introducing the new parameters 
$\rho _+ = r_+ ^2 \left/ \left( 2 \sqrt{r_\infty ^2 - r_+ ^2} \right) \right.$ and 
$\rho _0 = \left. \sqrt{r_\infty ^2 - r_+ ^2} \right/ 2$, 
then taking the limit $r_\infty \to \infty$, 
the equation \eqref{htemperature} represents the modified Hawking temperature 
of the five-dimensional Schwarzschild-Tangherlini black hole obtained by the uncharged fermion tunneling 
\cite{Feng:2015jlj}. 
Lastly, when $\rho _0 = 0 ,~ e = 0 ,~ \tilde L = 0$, 
the equation \eqref{htemperature} represents the modified Hawking temperature 
of the four-dimensional Reissner-Nordstr\"{o}m black hole by the uncharged fermion tunneling  
\cite{Chen:2013tha}.

\begin{figure}[!tbp]
\begin{center}
\includegraphics[scale=0.55]{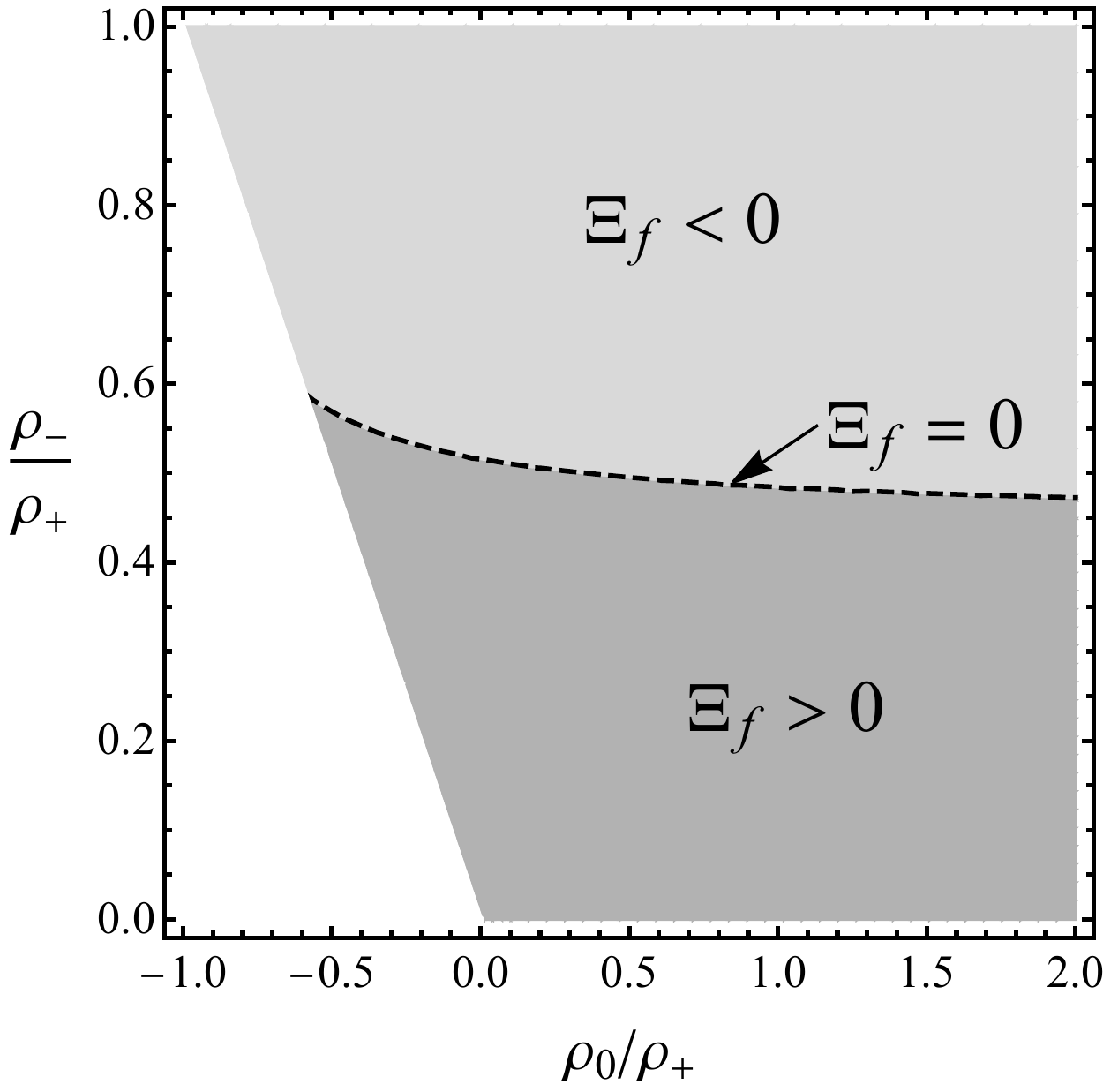} 
\caption{
Parameter regions of the correction \eqref{fcorrection} with $\rho_+ \geq \rho_- \geq 0 ,~ \rho_- + \rho_0 > 0$ 
for $m=0 ,~ \tilde L = 0 ,~ e / \tilde \omega = 0.1$.  
The positive $\Xi _f$ region (dark region) increases with increasing $e / \tilde \omega$.  
}
\label{fig:fregion}
\end{center}
\end{figure}

The evaporation process of the squashed Kaluza-Klein black hole 
by the quantum tunneling radiation of fermions 
depends on the correction $\Xi _f$ in the modified temperature $\tilde T$.   
Since all particles emitted by the Hawking radiation near the horizon region are effectively massless  
\cite{Parikh:1999mf, Srinivasan:1998ty, Angheben:2005rm, Kerner:2007rr, Kerner:2008qv}, 
the mass of the emitted fermion is not taken into account in the following discussion, i.e., $m = 0$.         
Further we assume that the fermion has no momentum in the direction of the extra dimension  
\cite{Matsuno:2009nz}, i.e., $\tilde L = 0$.    
We show the parameter regions of $\Xi _f$ in the figure \ref{fig:fregion}. 
We see that the correction \eqref{fcorrection} and its parameter regions are similar to those 
in the tunneling of scalar particles discussed in the section \ref{sec:stunneling}. 
Then the equation \eqref{htemperature} with the positive $\Xi _f$ shows that 
the generalized uncertainty principle slows down the increase of 
the Hawking temperature due to the radiation 
and the evaporation may cease at the particular mass of the black hole,  
leading to the thermodynamic stable Planck mass remnant. 
From the equation \eqref{fcorrection} and the figure \ref{fig:fregion}, we find that, 
while the black holes with the large $Q$ would be completely evaporated, 
the effectively four-dimensional black holes with the small $\rho _0$ and the small $Q$ would 
turn into the remnants at the end of the evaporation process.

\section{Summary and discussion}
\label{sec:summary}

We study the Hawking radiation of charged scalar particles and charged fermions 
from the five-dimensional charged static squashed Kaluza-Klein black hole 
on the basis of the tunneling mechanism, including the quantum gravity effects predicted by  
the quadratic generalized uncertainty principle with the minimal measurable length.   
We derive the modified Hawking temperature with  
the corrections related to the energy of the emitted particle, 
the compact extra dimension, the Maxwell field and the generalized uncertainty principle  
in the squashed Kaluza-Klein black hole background.     
Some known Hawking temperatures in five and four-dimensional black hole spacetimes  
are obtained by taking limits in the modified temperature.

We consider the evaporation process of the five-dimensional squashed Kaluza-Klein black hole 
with the generalized uncertainty principle by the tunneling of particles 
as one of the quantum gravity effects in the Hawking radiation.     
We see that the generalized uncertainty principle may prevent  
the complete evaporation of the squashed Kaluza-Klein black hole.  
As the black hole mass decreases due to the radiation,  
the black hole with the negative heat capacity may undergo a phase transition 
to the one with the positive heat capacity. 
At the minimum mass of the black hole, 
both the Hawking temperature and the heat capacity may vanish. 
This implies that the black hole may not exchange its energy with the surrounding environment.  
Then the evaporation of the squashed Kaluza-Klein black hole may cease, 
leading to the thermodynamic stable remnant.  
Related to the corrections in the modified Hawking temperature,  
we see that, while the black holes with the large charges would evaporate completely,  
the effectively four-dimensional black holes with the small charges would 
turn into the remnants at the end of the evaporation process. 
Further we consider the sparsity of the Hawking radiation from the squashed Kaluza-Klein black hole 
in the presence of the quantum gravity effect. 
We find that the sparsity may increase with decreasing the black hole mass 
and become infinite at the final stage of the evaporation.  
This also indicates that the generalized uncertainty principle 
may stop the Hawking radiation and lead to the black hole remnant.  
We see that the mass of the squashed Kaluza-Klein black hole remnant  
increases with increasing the deformation parameter of the generalized uncertainty principle, 
while decreases with increasing the size of the extra dimension.  
If the deformation parameter is of order $1$ 
and the extra dimension size is of order $0.1$ mm,   
the black hole remnant mass is of order the Planck mass.   
Since such a Planck mass remnant would be a ground state mass of the black hole, 
the squashed Kaluza-Klein black hole remnant would have no hair.  
This would be a desirable property as a dark matter 
\cite{Casadio:2017sze, Kuntz:2019gka, MacGibbon:1987my, Chen:2002tu, Chen:2004ft, DiGennaro:2021vev}. 
If the squashed Kaluza-Klein metric would describe the geometry around a primordial black hole and 
the generalized uncertainty principle considered in this paper would play an important role 
in the quantum nature of the black hole, 
the squashed Kaluza-Klein black hole remnants would be a dark matter candidate. 
Moreover, the variations of the deformation parameter and the extra dimension size 
provide specific signatures on the quantum features of 
the squashed Kaluza-Klein black hole solutions with the generalized uncertainty principle 
which would open the possibility of testing such higher-dimensional models 
by using future astronomical and astrophysical observations.

At present, gravity has been experimentally verified at $9 \times 10^{-5}$ kg or more  
\cite{Westphal:2020okx}
and the quantum control has been realized at $5 \times 10^{-11}$ kg or less 
\cite{2015PhRvA..92f1801U}. 
Then it is expected that theoretical and experimental studies of macroscopic quantum phenomena 
on the Planck mass scale in gravitational fields would bring significant progress in quantum gravity researches.
Since the energy of the higher-dimensional black hole remnant 
would decrease to a detectable range in the Large Hadron Collider, 
such remnants would be observed in future accelerator experiments  
\cite{Koch:2005ks, Nayak:2009fv, Mureika:2011hg, Bellagamba:2012wz, Gingrich:2020jpl}. 
If higher-dimensional black holes would be created in an accelerator 
and we assume that the five-dimensional squashed Kaluza-Klein black hole solutions 
would describe geometries around such black holes, 
we expect that our present work would make a contribution to the verification of the Hawking radiation 
and the extra dimension in asymptotically Kaluza-Klein spacetimes.

In this paper, we consider the Hawking radiation with the quantum gravity effects 
inspired from the modification on the commutation relation of the matter field 
in the fixed Kaluza-Klein background geometry. 
When the black hole mass approaches the order of the Planck mass due to the radiation, 
it would be expected that some quantum gravity effects would lead to some quantum fluctuations 
in the background metric. 
In four dimensions, such modified background geometries would be given by 
the quantum deformed Schwarzschild black holes 
\cite{Kazakov:1993ha} 
and the black holes in the noncommutative model 
\cite{Nicolini:2005vd} 
and the asymptotically safe gravity 
\cite{Bonanno:2000ep}. 
Since these black hole solutions are analogous to the Reissner-Nordstr\"{o}m solution,  
the quantum corrections in these frameworks affect things in the same way as the black hole charge  
\cite{Bonanno:2000ep, Nozari:2006rt, Bufalo:2014bwa, Hajebrahimi:2020xvo}.   
If there would exist a correspondence between the charged Kaluza-Klein black holes with squashed horizons 
and some quantum-corrected black holes in five dimensions, 
the black hole charge in the squashed Kaluza-Klein spacetime 
might have something to do with the very structure of the spacetime manifold and 
result in some quantum fluctuations in the background geometry as a quantum gravity effect. 
Moreover, if we take into account the effect of the emitted particle's self-gravitation 
in the Hawking evaporation process, 
the mass and the charge of the black hole may decrease to satisfy the energy conservation. 
Then the background metric may become dynamical by the backreaction effect of the quantum tunneling radiation 
\cite{Parikh:1999mf, Vanzo:2011wq, Nozari:2012nf}. 
We might regard such a dynamical geometry associated with the radiation 
as a modified background spacetime with some quantum fluctuations. 
Studies of these topics are currently in progress.

\section*{Acknowledgments}
The author would like to thank Dr. Hideki Ishihara, Dr. Yoshiyuki Morisawa, Dr. Ken-ichi Nakao, Dr. Hirotaka Yoshino and 
colleagues at the theoretical astrophysics and gravity group in Osaka City University 
for valuable suggestions and discussions.  
This work was partly supported by Osaka City University Advanced Mathematical Institute 
(MEXT Joint Usage/Research Center on Mathematics and Theoretical Physics JPMXP0619217849).


\begin{thebibliography}{999}

\bibitem{Hawking:1974sw}
S.~W.~Hawking,
%``Particle Creation by Black Holes,''
Commun. Math. Phys. \textbf{43}, 199-220 (1975); 
\textbf{46}, 206(E) (1976). 

\bibitem{Parikh:1999mf}
M.~K.~Parikh and F.~Wilczek,
%``Hawking radiation as tunneling,''
Phys. Rev. Lett. \textbf{85}, 5042-5045 (2000). 

\bibitem{Srinivasan:1998ty}
K.~Srinivasan and T.~Padmanabhan,
%``Particle production and complex path analysis,''
Phys. Rev. D \textbf{60}, 024007 (1999). 

\bibitem{Angheben:2005rm}
M.~Angheben, M.~Nadalini, L.~Vanzo and S.~Zerbini,
%``Hawking radiation as tunneling for extremal and rotating black holes,''
JHEP \textbf{05}, 014 (2005).  

\bibitem{Kerner:2007rr}
R.~Kerner and R.~B.~Mann,
%``Fermions tunnelling from black holes,''
Class. Quant. Grav. \textbf{25}, 095014 (2008). 

\bibitem{Kerner:2008qv}
R.~Kerner and R.~B.~Mann,
%``Charged Fermions Tunnelling from Kerr-Newman Black Holes,''
Phys. Lett. B \textbf{665}, 277-283 (2008). 

\bibitem{Vanzo:2011wq}
L.~Vanzo, G.~Acquaviva and R.~Di Criscienzo,
%``Tunnelling Methods and Hawking's radiation: achievements and prospects,''
Class. Quant. Grav. \textbf{28}, 183001 (2011).  


\bibitem{Kempf:1994su}
A.~Kempf, G.~Mangano and R.~B.~Mann,
%``Hilbert space representation of the minimal length uncertainty relation,''
Phys. Rev. D \textbf{52}, 1108-1118 (1995).

\bibitem{Kempf:1996nk}
A.~Kempf and G.~Mangano,
%``Minimal length uncertainty relation and ultraviolet regularization,''
Phys. Rev. D \textbf{55}, 7909-7920 (1997).

\bibitem{Hossenfelder:2012jw}
S.~Hossenfelder,
%``Minimal Length Scale Scenarios for Quantum Gravity,''
Living Rev. Rel. \textbf{16}, 2 (2013).


\bibitem{Isi:2013cxa}
M.~Isi, J.~Mureika and P.~Nicolini,
%``Self-Completeness and the Generalized Uncertainty Principle,''
JHEP \textbf{11}, 139 (2013). 


\bibitem{Adler:2001vs}
R.~J.~Adler, P.~Chen and D.~I.~Santiago,
%``The Generalized uncertainty principle and black hole remnants,''
Gen. Rel. Grav. \textbf{33}, 2101-2108 (2001).  

\bibitem{Hossenfelder:2003jz}
S.~Hossenfelder, M.~Bleicher, S.~Hofmann, J.~Ruppert, S.~Scherer and H.~Stoecker,
%``Collider signatures in the Planck regime,''
Phys. Lett. B \textbf{575}, 85-99 (2003).


\bibitem{Medved:2004yu}
A.~J.~M.~Medved and E.~C.~Vagenas,
%``When conceptual worlds collide: The GUP and the BH entropy,''
Phys. Rev. D \textbf{70}, 124021 (2004). 


\bibitem{Das:2008kaa}
S.~Das and E.~C.~Vagenas,
%``Universality of Quantum Gravity Corrections,''
Phys. Rev. Lett. \textbf{101}, 221301 (2008).   


\bibitem{Tawfik:2014zca}
A.~N.~Tawfik and A.~M.~Diab,
%``Generalized Uncertainty Principle: Approaches and Applications,''
Int. J. Mod. Phys. D \textbf{23}, no.12, 1430025 (2014).  


\bibitem{Petruzziello:2020wkd}
L.~Petruzziello and F.~Illuminati,
%``Quantum gravitational decoherence from fluctuating minimal length and deformation parameter at the Planck scale,''
Nature Commun. \textbf{12}, no.1, 4449 (2021).


\bibitem{Chen:2013pra}
D.~Chen, H.~Wu and H.~Yang,
%``Fermion's tunnelling with effects of quantum gravity,''
Adv. High Energy Phys. \textbf{2013}, 432412 (2013). 

\bibitem{Chen:2013ssa}
D.~Y.~Chen, Q.~Q.~Jiang, P.~Wang and H.~Yang,
%``Remnants, fermions` tunnelling and effects of quantum gravity,''
JHEP \textbf{11}, 176 (2013).  

\bibitem{Chen:2013tha}
D.~Chen, H.~Wu and H.~Yang,
%``Observing remnants by fermions' tunneling,''
JCAP \textbf{03}, 036 (2014).  

\bibitem{Chen:2014xgj}
D.~Chen, H.~Wu, H.~Yang and S.~Yang,
%``Effects of quantum gravity on black holes,''
Int. J. Mod. Phys. A \textbf{29}, no.26, 1430054 (2014).  


\bibitem{Feng:2015jlj}
Z.~W.~Feng, H.~L.~Li, X.~T.~Zu and S.~Z.~Yang,
%``Corrections to the thermodynamics of Schwarzschild-Tangherlini black hole and the generalized uncertainty principle,''
Eur. Phys. J. C \textbf{76}, no.4, 212 (2016). 


\bibitem{Casadio:2017sze}
R.~Casadio, P.~Nicolini and R.~da Rocha,
%``Generalised uncertainty principle Hawking fermions from minimally geometric deformed black holes,''
Class. Quant. Grav. \textbf{35}, no.18, 185001 (2018). 


\bibitem{Kuntz:2019gka}
I.~Kuntz and R.~Da Rocha,
%``GUP black hole remnants in quadratic gravity,''
Eur. Phys. J. C \textbf{80}, no.5, 478 (2020). 


\bibitem{Dobiasch:1981vh}
P.~Dobiasch and D.~Maison,
%``Stationary, Spherically Symmetric Solutions of Jordan's Unified Theory of Gravity and Electromagnetism,''
Gen. Relativ. Gravit. \textbf{14}, 231 (1982). 

\bibitem{Gibbons:1985ac}
G.~W.~Gibbons and D.~L.~Wiltshire,
%``Black Holes in Kaluza-Klein Theory,''
Ann. Phys. (N.Y.) \textbf{167}, 201 (1986); \textbf{176}, 393(E) (1987). 

\bibitem{Ishihara:2005dp}
H.~Ishihara and K.~Matsuno,
%``Kaluza-Klein black holes with squashed horizons,''
Prog. Theor. Phys. \textbf{116}, 417 (2006). 

\bibitem{Stelea:2008tt}
C.~Stelea, K.~Schleich, and D.~Witt,
%``On squashed black holes in Godel universes,''
Phys. Rev. D \textbf{78}, 124006 (2008).

\bibitem{Tomizawa:2012nk}
S.~Tomizawa and S.~Mizoguchi,
%``General Kaluza-Klein black holes with all six independent charges in five-dimensional minimal supergravity,''
Phys. Rev. D \textbf{87}, 024027 (2013).

\bibitem{Cai:2006td}
R.~G.~Cai, L.~M.~Cao and N.~Ohta,
%``Mass and thermodynamics of Kaluza-Klein black holes with squashed horizons,''
Phys. Lett. B \textbf{639}, 354-361 (2006). 

\bibitem{Kurita:2007hu}
Y.~Kurita and H.~Ishihara,
%``Mass and free energy in thermodynamics of squashed Kaluza-Klein black holes,''
Class. Quant. Grav. \textbf{24}, 4525-4532 (2007). 

\bibitem{Kurita:2008mj}
Y.~Kurita and H.~Ishihara,
%``Thermodynamics of Squashed Kaluza-Klein Black Holes and Black Strings -- A Comparison of Reference Backgrounds --,''
Class. Quant. Grav. \textbf{25}, 085006 (2008).

\bibitem{Ishihara:2007ni}
H.~Ishihara and J.~Soda,
%``Hawking radiation from squashed Kaluza-Klein black holes: A Window to extra dimensions,''
Phys. Rev. D \textbf{76}, 064022 (2007).


\bibitem{Matsuno:2011ca}
K.~Matsuno and K.~Umetsu,
%``Hawking radiation as tunneling from squashed Kaluza-Klein black hole,''
Phys. Rev. D \textbf{83}, 064016 (2011). 

\bibitem{Li:2011zzm}
H.~L.~Li,
%``Fermion tunneling from squashed black holes in the Goedel universe and charged Kaluza-Klein space-time,''
Chin. Phys. B \textbf{20}, 030402 (2011).


\bibitem{Stetsko:2014dda}
M.~M.~Stetsko,
%``Tunnelling of scalar and Dirac particles from squashed charged rotating Kaluza\textendash{}Klein black holes,''
Eur. Phys. J. C \textbf{76}, 48 (2016).


\bibitem{Kimura:2007cr}
M.~Kimura, K.~Murata, H.~Ishihara, and J.~Soda,
%``Stability of Squashed Kaluza-Klein Black Holes,''
Phys. Rev. D \textbf{77}, 064015 (2008); \textbf{96}, 089902(E) (2017). 

\bibitem{Kimura:2018whv}
M.~Kimura and T.~Tanaka,
%``Stability analysis of black holes by the $S$-deformation method for coupled systems,''
Classical Quantum Gravity \textbf{36}, 055005 (2019). 

\bibitem{Matsuno:2009nz}
K.~Matsuno and H.~Ishihara,
%``Geodetic Precession in Squashed Kaluza-Klein Black Hole Spacetimes,''
Phys. Rev. D \textbf{80}, 104037 (2009). 

\bibitem{Azreg-Ainou:2019ylk}
M.~Azreg-A\"\i{}nou, M.~Jamil, and K.~Lin,
%``Gyroscope precession frequency analysis of a five dimensional charged rotating Kaluza-Klein black hole,''
Chin. Phys. C \textbf{44}, 065101 (2020). 

\bibitem{Chen:2011wb}
S.~Chen and J.~Jing,
%``Thin accretion disk around a Kaluza\textendash{}Klein black hole with squashed horizons,''
Phys. Lett. B \textbf{704}, 641 (2011).

\bibitem{Zhu:2020cfn}
J.~Zhu, A.~B.~Abdikamalov, D.~Ayzenberg, M.~Azreg-Ainou, C.~Bambi, M.~Jamil, S.~Nampalliwar, A.~Tripathi, and M.~Zhou,
%``X-ray reflection spectroscopy with Kaluza-Klein black holes,''
Eur. Phys. J. C \textbf{80}, 622 (2020).  

\bibitem{Matsuno:2020kju}
K.~Matsuno,
%``Light deflection by squashed Kaluza-Klein black holes in a plasma medium,''
Phys. Rev. D \textbf{103}, no.4, 044008 (2021).

\bibitem{Liu:2010wh}
Y.~Liu, S.~Chen, and J.~Jing,
%``Strong gravitational lensing in a squashed Kaluza-Klein black hole spacetime,''
Phys. Rev. D \textbf{81}, 124017 (2010). 

\bibitem{Chen:2011ef}
S.~Chen, Y.~Liu, and J.~Jing,
%``Strong gravitational lensing in a squashed Kaluza-Klein Godel black hole,''
Phys. Rev. D \textbf{83}, 124019 (2011). 

\bibitem{Sadeghi:2012bj}
J.~Sadeghi, A.~Banijamali, and H.~Vaez,
%``Strong Gravitational Lensing in a Charged Squashed Kaluza- Klein Black hole,''
Astrophys. Space Sci. \textbf{343}, 559 (2013). 

\bibitem{Sadeghi:2013ssa}
J.~Sadeghi, J.~Naji, and H.~Vaez,
%``Strong gravitational lensing in a charged squashed Kaluza-Klein G\"odel black hole,''
Phys. Lett. B \textbf{728}, 170 (2014). 

\bibitem{Ji:2013xua}
L.~Ji, S.~Chen, and J.~Jing,
%``Strong gravitational lensing in a rotating Kaluza-Klein black hole with squashed horizons,''
J. High Energy Phys. 03 (2014) 089. 

\bibitem{Long:2019nox}
F.~Long, J.~Wang, S.~Chen, and J.~Jing,
%``Shadow of a rotating squashed Kaluza-Klein black hole,''
J. High Energy Phys. 10 (2019) 269. 

\bibitem{1828181}
M.~Ghasemi-Nodehi, M.~Azreg-A\"\i{}nou, K.~Jusufi, and M.~Jamil,
%``Shadow, quasinormal modes, and quasiperiodic oscillations of rotating Kaluza-Klein black holes,''
Phys. Rev. D \textbf{102}, 104032 (2020). 


\bibitem{Sasaki:2018dmp}
M.~Sasaki, T.~Suyama, T.~Tanaka and S.~Yokoyama,
%``Primordial black holes\textemdash{}perspectives in gravitational wave astronomy,''
Class. Quant. Grav. \textbf{35}, no.6, 063001 (2018). 


\bibitem{MacGibbon:1987my}
J.~H.~MacGibbon,
%``Can Planck-mass relics of evaporating black holes close the universe?,''
Nature \textbf{329}, 308-309 (1987). 

\bibitem{Chen:2002tu}
P.~Chen and R.~J.~Adler,
%``Black hole remnants and dark matter,''
Nucl. Phys. B Proc. Suppl. \textbf{124}, 103-106 (2003).  

\bibitem{Chen:2004ft}
P.~Chen,
%``Inflation induced Planck-size black hole remnants as dark matter,''
New Astron. Rev. \textbf{49}, 233-239 (2005).

\bibitem{Scardigli:2010gm}
F.~Scardigli, C.~Gruber and P.~Chen,
%``Black Hole Remnants in the Early Universe,''
Phys. Rev. D \textbf{83}, 063507 (2011).  

\bibitem{Dalianis:2019asr}
I.~Dalianis and G.~Tringas,
%``Primordial black hole remnants as dark matter produced in thermal, matter, and runaway-quintessence postinflationary scenarios,''
Phys. Rev. D \textbf{100}, no.8, 083512 (2019).  

\bibitem{Lehmann:2019zgt}
B.~V.~Lehmann, C.~Johnson, S.~Profumo and T.~Schwemberger,
%``Direct detection of primordial black hole relics as dark matter,''
JCAP \textbf{10}, 046 (2019).

\bibitem{Bai:2019zcd}
Y.~Bai and N.~Orlofsky,
%``Primordial Extremal Black Holes as Dark Matter,''
Phys. Rev. D \textbf{101}, no.5, 055006 (2020). 


\bibitem{Harris:2003eg}
C.~M.~Harris and P.~Kanti,
%``Hawking radiation from a (4+n)-dimensional black hole: Exact results for the Schwarzschild phase,''
JHEP \textbf{10}, 014 (2003).

\bibitem{Sampaio:2009tp}
M.~O.~P.~Sampaio,
%``Distributions of charged massive scalars and fermions from evaporating higher-dimensional black holes,''
JHEP \textbf{02}, 042 (2010).

\bibitem{Arbey:2021yke}
A.~Arbey, J.~Auffinger, M.~Geiller, E.~R.~Livine and F.~Sartini,
%``Hawking radiation by spherically-symmetric static black holes for all spins. II. Numerical emission rates, analytical limits, and new constraints,''
Phys. Rev. D \textbf{104}, no.8, 084016 (2021).


\bibitem{Bonanno:2000ep}
A.~Bonanno and M.~Reuter,
%``Renormalization group improved black hole space-times,''
Phys. Rev. D \textbf{62}, 043008 (2000).

\bibitem{Myung:2006mz}
Y.~S.~Myung, Y.~W.~Kim and Y.~J.~Park,
%``Thermodynamics and evaporation of the noncommutative black hole,''
JHEP \textbf{02}, 012 (2007).

\bibitem{DiGennaro:2021vev}
S.~Di Gennaro and Y.~C.~Ong,
%``Feasibility of primordial black hole Remnants as dark matter in view of Hawking radiation recoil,''
JCAP \textbf{07}, 041 (2021). 


\bibitem{Page:1976df}
D.~N.~Page,
%``Particle Emission Rates from a Black Hole: Massless Particles from an Uncharged, Nonrotating Hole,''
Phys. Rev. D \textbf{13}, 198-206 (1976). 

\bibitem{Gray:2015pma}
F.~Gray, S.~Schuster, A.~Van-Brunt and M.~Visser,
%``The Hawking cascade from a black hole is extremely sparse,''
Class. Quant. Grav. \textbf{33}, no.11, 115003 (2016).

\bibitem{Hod:2016rmg}
S.~Hod,
%``The Hawking cascades of gravitons from higher-dimensional Schwarzschild black holes,''
Phys. Lett. B \textbf{756}, 133-136 (2016).

\bibitem{Paul:2016xvb}
A.~Paul and B.~R.~Majhi,
%``Hawking evaporation cascade in the presence of backreaction effect,''
Int. J. Mod. Phys. A \textbf{32}, no.16, 1750088 (2017).

\bibitem{Miao:2017jtr}
Y.~G.~Miao and Z.~M.~Xu,
%``Hawking Radiation of Five-dimensional Charged Black Holes with Scalar Fields,''
Phys. Lett. B \textbf{772}, 542-546 (2017).

\bibitem{Alonso-Serrano:2018ycq}
A.~Alonso-Serrano, M.~P.~Dabrowski and H.~Gohar,
%``Generalized uncertainty principle impact onto the black holes information flux and the sparsity of Hawking radiation,''
Phys. Rev. D \textbf{97}, no.4, 044029 (2018).  

\bibitem{Ong:2018syk}
Y.~C.~Ong,
%``An effective black hole remnant via infinite evaporation time due to generalized uncertainty principle,''
JHEP \textbf{10}, 195 (2018). 

\bibitem{Feng:2018jqf}
Z.~W.~Feng, X.~Zhou, S.~Q.~Zhou and D.~D.~Feng,
%``Rainbow gravity corrections to the information flux of a black hole and the sparsity of Hawking radiation,''
Annals Phys. \textbf{416}, 168144 (2020).

\bibitem{Alonso-Serrano:2020hpb}
A.~Alonso-Serrano, M.~P.~Dabrowski and H.~Gohar,
%``Nonextensive Black Hole Entropy and Quantum Gravity Effects at the Last Stages of Evaporation,''
Phys. Rev. D \textbf{103}, no.2, 026021 (2021).

\bibitem{Schuster:2019xvp}
S.~Schuster,
%``Sparsity of Hawking radiation in $D$ + 1 space-time dimensions for massless and massive particles,''
Class. Quant. Grav. \textbf{38}, no.4, 047002 (2021).


\bibitem{Westphal:2020okx}
T.~Westphal, H.~Hepach, J.~Pfaff and M.~Aspelmeyer,
%``Measurement of gravitational coupling between millimetre-sized masses,''
Nature \textbf{591}, no.7849, 225-228 (2021).

\bibitem{2015PhRvA..92f1801U} 
M.~Underwood, D.~Mason, D.~Lee, H.~Xu, L.~Jiang, A.~B.~Shkarin, K.~B{\o}rkje, S.~M. Girvin, and J.~G.~E. Harris, 
Phys. Rev. A \textbf{92}, 061801(R) (2015). 


\bibitem{Koch:2005ks}
B.~Koch, M.~Bleicher and S.~Hossenfelder,
%``Black hole remnants at the LHC,''
JHEP \textbf{10}, 053 (2005). 


\bibitem{Nayak:2009fv}
G.~C.~Nayak,
%``Dark Matter Production at LHC from Black Hole Remnants,''
Phys. Part. Nucl. Lett. \textbf{8}, 337-341 (2011).  

\bibitem{Mureika:2011hg}
J.~Mureika, P.~Nicolini and E.~Spallucci,
%``Could any black holes be produced at the LHC?,''
Phys. Rev. D \textbf{85}, 106007 (2012). 

\bibitem{Bellagamba:2012wz}
L.~Bellagamba, R.~Casadio, R.~Di Sipio and V.~Viventi,
%``Black Hole Remnants at the LHC,''
Eur. Phys. J. C \textbf{72}, 1957 (2012). 

\bibitem{Gingrich:2020jpl}
D.~M.~Gingrich and B.~Undseth,
%``Quantum black holes in the horizon quantum mechanics model at the Large Hadron Collider,''
Phys. Rev. D \textbf{102}, no.9, 095020 (2020).  


\bibitem{Kazakov:1993ha}
D.~I.~Kazakov and S.~N.~Solodukhin,
%``On Quantum deformation of the Schwarzschild solution,''
Nucl. Phys. B \textbf{429}, 153-176 (1994). 

\bibitem{Nicolini:2005vd}
P.~Nicolini, A.~Smailagic and E.~Spallucci,
%``Noncommutative geometry inspired Schwarzschild black hole,''
Phys. Lett. B \textbf{632}, 547-551 (2006).


\bibitem{Nozari:2006rt}
K.~Nozari and B.~Fazlpour,
%``Reissner-Nordstrom Black Hole Thermodynamics in Noncommutative Spaces,''
Acta Phys. Polon. B \textbf{39}, 1363-1374 (2008).

\bibitem{Bufalo:2014bwa}
R.~Bufalo and A.~Tureanu,
%``Analogy between the Schwarzschild solution in a noncommutative gauge theory and the Reissner-Nordstr\"om metric,''
Phys. Rev. D \textbf{92}, no.6, 065017 (2015).

\bibitem{Hajebrahimi:2020xvo}
M.~Hajebrahimi and K.~Nozari,
%``A quantum-corrected approach to black hole radiation via a tunneling process,''
PTEP \textbf{2020}, no.4, 043E03. 


\bibitem{Nozari:2012nf}
K.~Nozari and S.~Saghafi,
%``Natural Cutoffs and Quantum Tunneling from Black Hole Horizon,''
JHEP \textbf{11}, 005 (2012).


\end{thebibliography}
\end{document}